\documentclass[aps,twocolumn,prl,superscriptaddress,amsmath,showpacs,tightenlines]{revtex4-1}
\usepackage{amssymb}
\usepackage{amsmath}
\usepackage{braket}
\usepackage{graphicx}
\usepackage{epsfig}
\usepackage{subfigure}
\usepackage{amsfonts}
\usepackage{CJK}
\usepackage{appendix}
\usepackage{multirow}
\usepackage{amsfonts}
\usepackage{dsfont}
\usepackage{amsmath}
\usepackage{graphicx}
\usepackage{float}
\usepackage{amsmath,epsfig,amssymb,subfigure,bm,dsfont}
\usepackage{lipsum}
\usepackage[english]{babel}
\usepackage{bm}
\usepackage{amsmath}

\begin{document}
\title{Bound state in the continuum and multiple atom state transfer applications in a waveguide QED setup}
\author{Xiang Guo}
\affiliation{Center for Quantum Sciences and School of Physics, Northeast Normal University, Changchun 130024, China}
\author{Xiaojun Zhang}
\affiliation{Center for Quantum Sciences and School of Physics, Northeast Normal University,
Changchun 130024, China}
\author{Mingzhu Weng}
\affiliation{School of Physics Science and Technology, Shenyang Normal University, Shenyang 110034, China}
\author{Qian Bin}
\affiliation{College of Physics, Sichuan University, Chengdu 610065, China}
\author{Hao-di Liu}
\affiliation{Center for Quantum Sciences and School of Physics, Northeast Normal University,
Changchun 130024, China}
\author{Hai-Jun Xing}
\email{hjxing@nenu.edu.cn}
\affiliation{Center for Quantum Sciences and School of Physics, Northeast Normal University,
Changchun 130024, China}
\author{Xin-You L\"{u}}
\email{xinyoulu@hust.edu.cn}
\affiliation{School of Physics and Institute for Quantum Science and Engineering, Huazhong University of Science and Technology, and Wuhan Institute of Quantum Technology, Wuhan 430074, China}
\author{Zhihai Wang}
\email{wangzh761@nenu.edu.cn}
\affiliation{Center for Quantum Sciences and School of Physics, Northeast Normal University,
Changchun 130024, China}\
\begin{abstract}
Bound states in the continuum (BICs) have been extensively exploited to enhance light--matter interactions in metamaterials, yet their emergence and utility in multi-atom waveguide platforms remain far less explored. Here we study atom--waveguide-dressed BICs in a one-dimensional coupled-resonator waveguide, where two spatially separated atomic arrays couple to distinct resonators with time-dependent strengths. We show that these BICs support a standing-wave photonic mode and enable the transfer of an arbitrary unknown quantum state between the two arrays with fidelities exceeding $99\%$. The protocol remains robust against both disorder and intrinsic dissipation. Our results establish BICs as long-lived resources for high-fidelity quantum information processing in waveguide-QED architectures.
\end{abstract}

\maketitle

\emph{Introduction}-Waveguide QED~\cite{DR2017,DE2018,AS2023_1} investigates strong light--matter interactions in confined photonic structures and explores their applications in quantum networks~\cite{XG2017,IC2020,NG2020,AB2021} and quantum information processing~\cite{JC2008,GW2017,AB2020,MK2020}. A key resource in waveguide-QED systems is the atom--photon bound state outside the continuum (BOC)~\cite{TS2009,GC2016,YL2017,LK2018,WZ2020,AS2023_2}, which has been exploited to preserve coherence~\cite{CJ2016}, protect entanglement~\cite{QJ2010}, and enhance quantum-precision measurements~\cite{AW2012,KB2019}. More recently, it has been shown that waveguide-QED platforms can also support bound states in the continuum (BICs)~\cite{FH2019,DC2008,MI2012}.

BICs have been extensively studied in the metamaterials community as a route to enhancing light--matter interactions~\cite{SI2021,QQ2023,MK2023}. By contrast, their counterparts in waveguide-QED platforms remain much less explored. In a waveguide, the photonic component of a BIC forms a standing-wave pattern pinned by the atoms, rather than exhibiting the exponential spatial localization characteristic of bound states outside the continuum (BOCs)~\cite{WZ2020,XZ2023w}. At the same time, BICs inherit the long lifetime associated with BOCs. Therefore, it can mediate effective atomic crosstalk over distances that exceed the BOC localization length. This observation motivates us to investigate BIC-enabled quantum information processing between remote atomic nodes. A representative application is high-fidelity quantum state transfer~\cite{JI1997,KI2016,ZL2017,PK2018,CJ2018,PM2020,BV2017,WK2020,YX2025,ZL2023}.

In this Letter, we investigate BICs in a waveguide-QED system where two arrays of two-level atoms couple to a coupled-resonator waveguide (CRW)~\cite{MJ2006,MN2008,XZ2023} at two spatially separated sites with time-modulated coupling strengths. We find that the BIC eigenenergy (eigenfrequency) is time independent, while its wave function can be dynamically shaped through the controlled couplings. Moreover, we analytically show that these BICs can encode atomic entanglement and enable the transfer of \emph{arbitrary, unknown} entangled states between the two arrays. Our BIC-based state-transfer protocol requires neither long-distance pre-shared entanglement nor intermediate measurements---in sharp contrast to quantum teleportation~\cite{CH1993,DC2017,YH2019,BL2022}.

Ever since the pioneering state-transfer proposal based on cascaded quantum networks~\cite{JI1997}, a variety of strategies have been developed with the aid of nonreciprocal elements (e.g., unidirectional waveguides and port circulators)~\cite{KI2016,ZL2017,PK2018,CJ2018,PM2020} or chiral atom--waveguide coupling~\cite{BV2017,WK2020,AS2023_1,JP2014,RM2014,PL2017,DG2025}. However, these approaches are often vulnerable to insertion loss and network noise~\cite{NL2019,IS2015,SX2021,DH2022}. By contrast, our BIC-based state-transfer scheme requires neither nonreciprocal devices nor chiral couplings. For atomic arrays, fidelities above $99\%$ can be achieved in both adiabatic and nonadiabatic regimes, with the latter reducing the required time by roughly an order of magnitude. More importantly, benefiting from the standing-wave character of the BIC~\cite{XZ2023w} and its robustness against disorder~\cite{MW2025}, our protocol remains resilient to disorder in the waveguide on-site frequencies, inter-site hopping rates, and atom--waveguide couplings, as well as to intrinsic dissipation of both atoms and resonators~\cite{AP2021,SG2024,JB2024,EV2023,AE2025,YT2025}.

\emph {Model and bound state in the continuum}-As shown in Fig.~\ref{Model}, we consider two arrays of two-level atoms, each comprising $N_a$ atoms, coupled to the $N_1$th and $N_2$th resonators of a one-dimensional CRW. The CRW is described by the tight-binding Hamiltonian (hereafter $\hbar=1$)
\begin{figure}
  \centering
  \includegraphics[width=0.96\columnwidth]{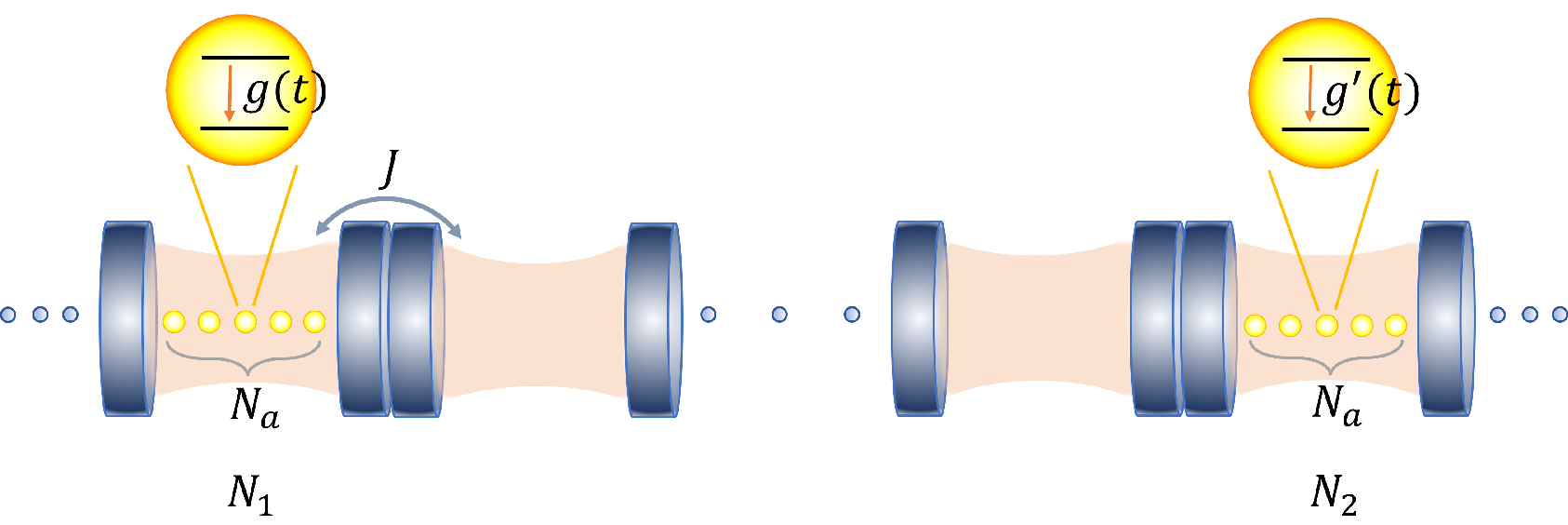}\\
  \caption{Schematic of BIC-enabled state transfer between two atomic arrays in a coupled-resonator waveguide. The waveguide consists of coupled resonators with hopping strength $J$. Each array contains $N_a$ atoms and couples to the $N_1$th and $N_2$th resonators with time-dependent strengths $g(t)$ and $g'(t)$, respectively.}
  \label{Model}
\end{figure}
\begin{equation}
\hat{H}_{c}=\sum_{n}\omega_{c}\hat{a}_{n}^{\dagger}\hat{a}_{n}
-J\sum_{n}(\hat{a}_{n+1}^{\dagger}\hat{a}_{n}+\text{H.c.}),
\label{CRW}
\end{equation}
where $\hat{a}_n$ is the bosonic annihilation operator in the $n$th resonator. The waveguide supports a continuous band of width $4J$ centered at $\omega_c$, with dispersion $\omega_k=\omega_{c}-2J\cos k$, where $k$ is the photon wave vector. In what follows, we set $\omega_c=0$ as the reference frequency.

The full system Hamiltonian reads
\begin{eqnarray}
\hat{H}(t)&=&\hat{H}_{c}+\sum_{i=1}^{N_{a}}\frac{\Omega_{i}}{2}
\left(\hat{\sigma}_{z}^{(i)}+\hat{\sigma}_{z}^{(N_a+i)}\right)\nonumber \\
&+&\sum_{i=1}^{N_{a}}\left[g(t)\hat{a}_{N_1}^{\dagger}\hat{\sigma}_{-}^{(i)}+g'(t)\hat{a}_{N_2}^{\dagger}
\hat{\sigma}_{-}^{(N_{a}+i)}+\text{H.c.}\right],
\label{H(t)}
\end{eqnarray}
where $\hat{\sigma}_{z,\pm}^{(i)}$ are Pauli operators for the $i$th atom with transition frequency $\Omega_i$.
Here $g(t)$ [$g'(t)$] denotes the time-dependent coupling between the $N_1$th [$N_2$th] resonator and the atoms
in the first (second) array. We emphasize that the $i$th atom
in the first array and the $(N_a+i)$th ($i=1,2,\ldots,N_a$) atom in the second array share a common frequency
$\Omega_i$, while $\Omega_i\neq\Omega_j$ for $i\neq j$. Under the rotating-wave approximation, the total excitation
number is conserved. In what follows, we restrict our analysis to the single-excitation subspace.

Since the atom-CRW couplings are time dependent, all eigenvalues of the Hamiltonian would in principle be time dependent as well. However, photons in the CRW are repeatedly reflected by the atoms at the $N_1$th and $N_2$th resonators, giving rise to strong interference effects. As shown in detail in the Supplemental Material (SM)~\cite{SM}, one can prove that the matrix
$\hat{O}(t)=\hat{H}(t)-\Omega_p \hat{I}$ is not full rank at all times provided that the condition
$K_p\Delta N = m_p \pi$ is satisfied for all $p=1,2,\dots,N_a$, where $m_p \in \mathbb{Z}^{+}$ and
$\Delta N = N_2 - N_1$ is the distance between the two atomic arrays. Here $K_p$ is the wave vector of a photon resonant with the $p$th and $(N_a+p)$th atoms, i.e., $-2J\cos K_p = \Omega_p$. This implies that the Hamiltonian supports a set of $N_a$ time-independent eigenvalues, $\Omega_p$ $(p=1,2,\dots,N_a)$, which are exactly resonant with the corresponding atomic transition frequencies. We denote the instantaneous eigenstate with eigenfrequency $\Omega_p$ as
\begin{align}
\ket{\psi^{(p)}(t)}
&=
\Biggl[
\sum_{n}\alpha_{n}^{(p)}(t)\hat{a}^{\dagger}_{n}
+\sum_{i=1}^{2N_a}\beta_{i}^{(p)}(t)\hat{\sigma}_{+}^{(i)}
\Biggr]\ket{G},
\label{distribution}
\end{align}
where $\ket{G}$ denotes the global ground state with all atoms in their ground state and the CRW in the vacuum. Substituting this ansatz into the Schr\"odinger equation
$\hat{H}(t)\ket{\psi^{(p)}(t)}=\Omega_p\ket{\psi^{(p)}(t)}$, we obtain
\begin{equation}
\beta_{p}^{(p)}(t)=\frac{f(t)}{g(t)},\qquad
\beta_{N_a+p}^{(p)}(t)=\frac{f(t)}{g'(t)},
\label{beta}
\end{equation}
and $\beta_i^{(p)}(t)=0$ for $i\neq p,N_a+p$. The time-dependent modulation factor is given by
$f(t)=1/\sqrt{[g(t)]^{-2}+[g'(t)]^{-2}+(\Delta N/2)\,[J\sin K_p]^{-2}}$,
which is invariant under exchanging $g(t)$ and $g'(t)$. The corresponding photonic component forms a standing wave localized between the two atomic arrays, with amplitudes
\begin{equation}
\alpha_{n}^{(p)}(t)
=f(t)\,\frac{\sin[(n-N_{1})K_{p}]}{J\sin K_p},
\label{alpha_p}
\end{equation}
for $N_1\leq n \leq N_2$, and $\alpha_n^{(p)}(t)=0$ otherwise.

The distinctive features of the instantaneous eigenstate $\ket{\psi^{(p)}(t)}$ can be summarized as follows.
First, as indicated by Eq.~\eqref{beta}, the atomic amplitudes are directly tunable via the couplings $g(t)$ and $g'(t)$.
The $p$th and $(N_a+p)$th atoms can thus exchange their excitation as one effectively interchanges $g(t)$ and $g'(t)$.
In particular, in the limit $g(t)\!\to\!0$ we obtain $\beta_{p}^{(p)}(t)\!\to\!1$ and $\beta_{N_a+p}^{(p)}(t)\!\to\!0$,
whereas in the limit $g'(t)\!\to\!0$ we have $\beta_{p}^{(p)}(t)\!\to\!0$ and $\beta_{N_a+p}^{(p)}(t)\!\to\!1$.
Second, the photonic component described by Eq.~\eqref{alpha_p} reveals that the two atomic arrays act as effective mirrors,
confine photons between them, and form a standing-wave pattern, i.e., a bound state in real space.
In this Letter, we focus on the parameter regime $\Omega_i\in(-2J,2J)$ for $i=1,2,\ldots,N_a$.
Because these states lie within the continuous band in frequency space, they are in fact BICs~\cite{FH2019,MI2012,DC2008}.
In what follows, we denote them as $\ket{{\rm BIC}^{(i)}}$ to distinguish them from other scattering states.

Throughout this Letter, we adopt a simple linear time-dependent modulation,
$g(t)=aJt/T, g'(t)=aJ(1-t/T)$, which satisfies $g(t)=g'(T-t)$, $g(0)/g'(0)\simeq 0$, and $g(T)/g'(T)\simeq \infty$.
Given current experimental capabilities, the CRW can be implemented with superconducting circuit architectures~\cite{XZ2023}.
We take the photon hopping strength to be $J=2\pi\times200$~MHz.
The atoms can be realized by artificial transmon qubits, and both the resonators and transmons operate in the microwave regime.
With these parameters, we plot in the SM~\cite{SM} (see Fig.~S1) the energy spectra of each BIC together with several nearby scattering states for $N_a=3$ and $N_a=4$.
In what follows, we denote the scattering state closest to the $p$th BIC as $\ket{{\rm SC}^{(p)}}$.

Including the resonator losses in the CRW, the system can be approximately described by the non-Hermitian Hamiltonian $\hat{H}'(t)=\hat{H}(t)-i\kappa/2\sum_{n}\hat{a}_n^\dagger \hat{a}_n$, where $\kappa$ is the decay rate of each resonator. The eigenfrequencies of $\hat{H}'(t)$ are therefore complex in general.
In Fig.~\ref{imaginary}(a), we plot the imaginary parts of the eigenfrequencies,
${\mathcal{K}}(t)=-{\rm Im}[E(t)]$,
which characterize the lifetimes of the corresponding modes in the system with $N_a=3$. As shown in the figure, all BICs exhibit the same and smallest values of ${\mathcal{K}}(t)$, indicating that they decay more slowly than other eigenmodes.
This behavior stems from the fact that these BICs confine photons primarily to the region between the two atomic arrays, so that photon loss outside this region does not significantly reduce their lifetimes.
Moreover, the nearest scattering states $\ket{{\rm SC}^{(1,2,3)}}$ also possess relatively small ${\mathcal{K}}(t)$.
By contrast, for scattering states far from the BICs, the photonic wave functions extend over the entire waveguide with substantial weight, leading to much larger imaginary parts of the eigenfrequencies and hence shorter lifetimes.

\begin{figure}
  \centering
  \includegraphics[width=0.48\columnwidth]{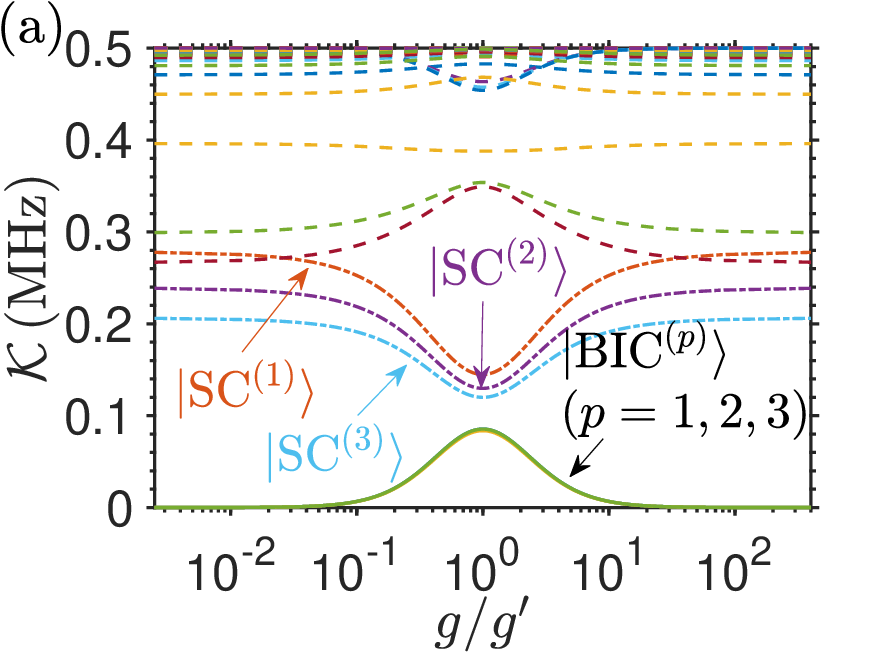}
  \includegraphics[width=0.48\columnwidth]{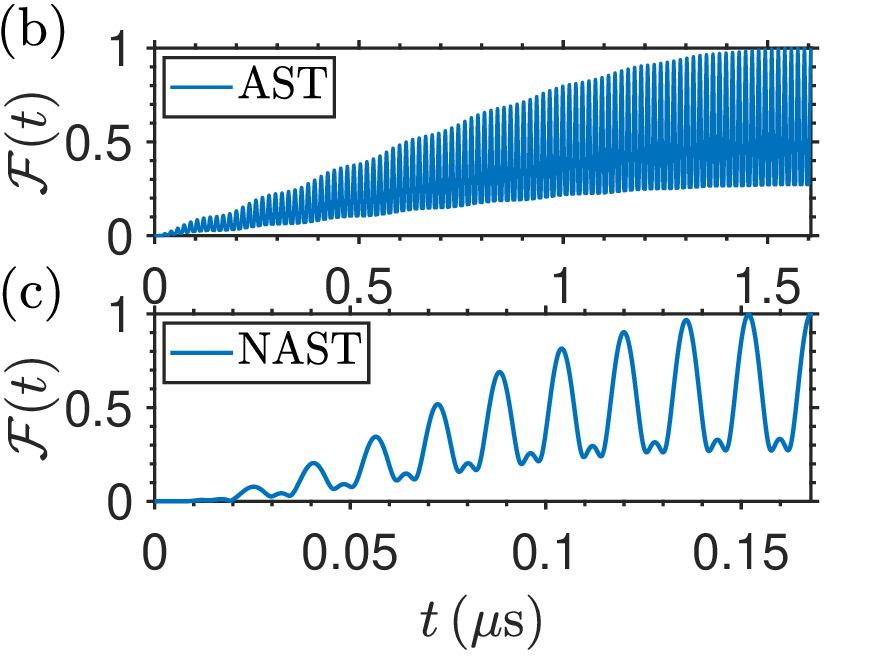}
  \caption{(a) Imaginary parts of the eigenfrequencies of $\hat{H}'(t)$ as a function of the coupling ratio $g/g'$. (b) and (c) Dynamics of the AST and NAST fidelities, respectively. We set $J=2\pi\times 200\,{\rm MHz}$, $a=0.4J$, $N_{c}=41$, $N_{a}=3$, $\Delta N=20$, $\Omega_{1}=-2J\cos(11\pi/20)$, $\Omega_{2}=0$, and $\Omega_{3}=-2J\cos(9\pi/20)$ in all panels. The remaining parameters are $T=1.6061\,{\rm \mu s}$ in (b) and $T=0.1678\,{\rm \mu s}$ in (c).}
  \label{imaginary}
\end{figure}

In waveguide-QED platforms, BICs originate from destructive interference and can manifest as a persistent but otherwise decaying oscillation~\cite{SL2021,KH2023,AS2023,ER2024,HY2025}.
For the two-atom CRW-coupling scenario, BICs have been shown to be robust against both on-site and hopping disorder and have been exploited for high-fidelity entangled-state preparation~\cite{MW2025}.
Here, guided by Eq.~\eqref{beta}, we show that BICs can encode an arbitrary atomic state, while the time-reversal-symmetric modulation of the couplings enables state transfer.
Moreover, the standing-wave character of the BIC provides a stable information channel, and its long lifetime ensures high fidelities even under nonideal conditions with atomic emission and photonic dissipation, as demonstrated below.
As a first step, we consider adiabatic state transfer (AST), in which the system dynamics remain within the BIC manifold, to transfer an arbitrary entangled state from the atoms coupled to the $N_1$th resonator to those coupled to the $N_2$th resonator.
To further shorten the transfer time, we then develop a nonadiabatic state-transfer (NAST) protocol that intentionally involves the nearest scattering states.

\emph{Adiabatic state transfer}-The energy spectrum in Fig.~S1 of the SM~\cite{SM} shows a finite gap between the $p$th BIC and its nearest scattering states. This separation suggests that atomic entangled states can be adiabatically transferred by staying within the BIC manifold.

Working in the single-excitation subspace, we aim to transfer an arbitrary state
\begin{equation}
\ket{\phi(0)}=\sum_{i=1}^{N_{a}}C_{i}\hat{\sigma}_{+}^{(i)}\ket{G},\,
\sum_{i=1}^{N_a}|C_i|^2=1.
\end{equation}
Since $g(0)\ll g'(0)$, the initial state can be encoded in the BIC manifold as
$\ket{\phi(0)}=\sum_{p=1}^{N_{a}}C_{p}\ket{{\rm BIC}^{(p)}(0)} $.
Under the adiabatic condition $|F_p(t)/\Delta_p(t)|\ll 1$, where
$F_p(t)=\bigl\langle {\rm BIC}^{(p)}(t)\big|\partial/\partial t\big|{\rm SC}^{(p)}(t)\bigr\rangle$ characterizes the $\hat{H}(t)$-induced transition rate between the $p$th BIC and its nearest scattering state, and $\Delta_p(t)=\Omega_p-E_S^{(p)}(t)$
denotes their instantaneous detuning (with $E_S^{(p)}(t)$ the eigenfrequency of $\ket{{\rm SC}^{(p)}(t)}$), the population in each BIC is preserved during the evolution. Consequently, at time $T$ the state becomes $\ket{\phi(T)}
=\sum_{p=1}^{N_{a}}C_{p}e^{-i\Omega_{p}T}\ket{{\rm BIC}^{(p)}(T)}
=\sum_{i=1}^{N_{a}}e^{-i\Omega_{i}T}(-1)^{m_{i}+1}C_{i}\,
\hat{\sigma}_{+}^{(N_a+i)}\ket{G}$. By choosing an appropriate $T$ such that
$e^{-i\Omega_{i}T}(-1)^{m_{i}+1}=e^{i\phi}$
for all $i=1,2,\ldots,N_a$, with a global phase $\phi$ independent of $i$,
we realize a perfect AST process.

In Fig.~\ref{imaginary}(b), we plot the fidelity $\mathcal{F}$ versus time for $N_a=3$.
The fidelity exhibits an oscillatory rise and reaches the maximal value of $100\%$ for a duration of $T\simeq 1.6\,\mu{\rm s}$.
Here we choose a representative initial state with
$C_{1}=\sqrt{2}/2$, $C_{2}=\sqrt{3}/3$, and $C_{3}=(\sqrt{6}/6)e^{i\pi/5}$;
the performance for an arbitrary input state is expected to be similar.
With current experimental capabilities, the artificial atoms can be implemented using superconducting transmon qubits,
whose coherence times are on the order of $10\,\mu{\rm s}$~\cite{MK2020}.
Thus, the time required for perfect AST is already comparable to the available coherence window. This observation motivates us to go beyond the adiabatic protocol and develop a faster state-transfer scheme,
as discussed below.

\emph{Non-adiabatic state transfer}-We now go beyond the AST process by allowing controlled tunneling from the BIC $\ket{{\rm BIC}^{(p)}(t)}$ to its nearest scattering state $\ket{{\rm SC}^{(p)}(t)}$ during the intermediate stage. Importantly, by the end of the protocol the population in the scattering state is nearly steered back into the BIC manifold, thereby guaranteeing a high transfer fidelity.

To describe the NAST process, we expand the quantum state at time $t$ as
\begin{align}
\ket{\phi(t)}=\sum_{p=1}^{N_{a}}C_{p}\Big[\tilde{C}_{B}^{(p)}(t)
\ket{{\rm BIC}^{(p)}(t)}+\tilde{C}_{S}^{(p)}(t)\ket{{\rm SC}^{(p)}(t)}\Big],
\end{align}
with the initial conditions $\tilde{C}_{B}^{(p)}(0)=1$ and $\tilde{C}_{S}^{(p)}(0)=0$ for $p=1,2,\cdots,N_a$.
The Schr\"odinger equation $i\partial \ket{\phi(t)}/\partial t=\hat{H}(t)\ket{\phi(t)}$ yields (see details in SM~\cite{SM})
\begin{align}
\dot{\tilde{C}}_{B}^{(p)}(t)+i\delta_{p}\tilde{C}_{B}^{(p)}(t)+F_p(t)\tilde{C}_{S}^{(p)}(t)=0,\nonumber\\
\dot{\tilde{C}}_{S}^{(p)}(t)+iE_{S}^{(p)}(t)\tilde{C}_{S}^{(p)}(t)-F_p(t)\tilde{C}_{B}^{(p)}(t)=0,
\label{Rabi}
\end{align}
which can be solved numerically. A central task is to choose an appropriate time $T$ such that
$(-1)^{m_{p}+1}\tilde{C}_{B}^{(p)}(T)\approx e^{i\Phi}$ and $\tilde{C}_{S}^{(p)}(T)\approx 0$
hold for all $p=1,2,\ldots,N_a$. This completes the transfer of an arbitrary state up to a global phase.
In Fig.~\ref{imaginary}(c), we plot the fidelity as a function of time for the NAST protocol. The maximum fidelity exceeds $99\%$,
while the required time is shortened by about one order of magnitude compared with the AST process, and is thus well below
the coherence time of superconducting transmon qubits, ensuring experimental feasibility.

In Fig.~\ref{non-adiabatic}(a), we further plot the photonic dynamics during the NAST protocol for $N_a=3$.
We find that the photon emitted by the atoms coupled to the $N_1$th resonator is largely confined to the region between the two atomic arrays, with only a small fraction leaking outside.
Equivalently, the photon is almost completely reflected by the second array as it propagates toward the $N_2$th resonator.
As a result, throughout the state-transfer process, the photon---as the information carrier---is effectively trapped by the BIC manifold together with the neighboring scattering states. This mechanism enables high-fidelity state transfer in a \emph{bidirectional} waveguide, going beyond most previous schemes that rely on nonreciprocal devices or chiral couplings as essential ingredients~\cite{KI2016,ZL2017,PK2018,CJ2018,PM2020,BV2017,WK2020}.

\begin{figure}
  \centering
  \includegraphics[width=0.48\columnwidth]{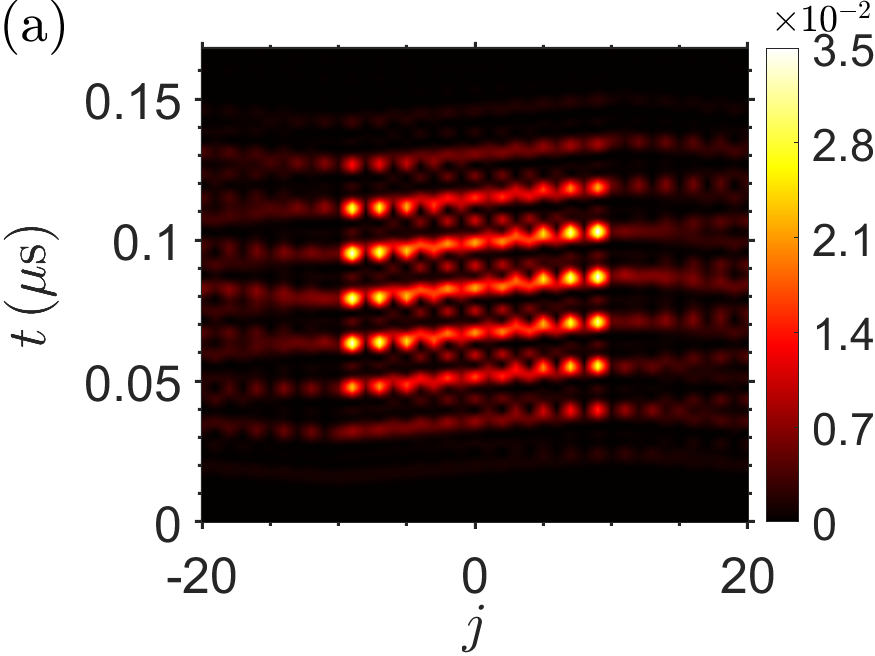}
  \includegraphics[width=0.48\columnwidth]{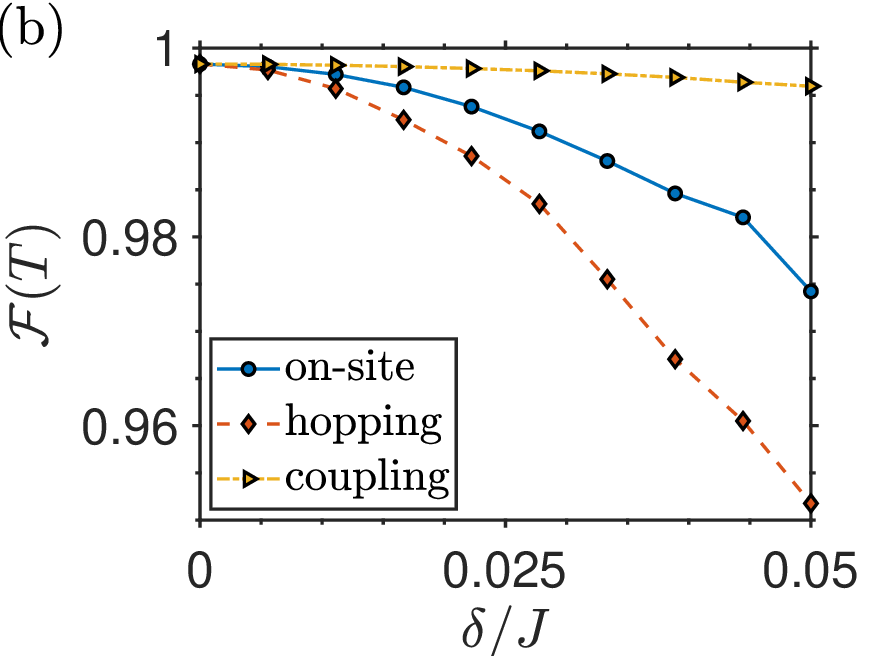}\\
  \includegraphics[width=0.48\columnwidth]{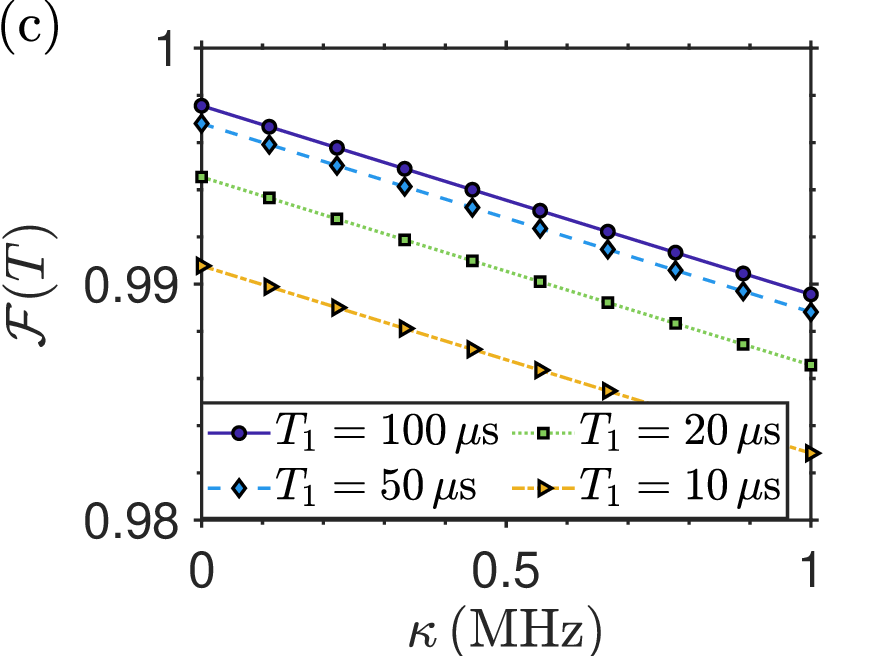}
  \includegraphics[width=0.48\columnwidth]{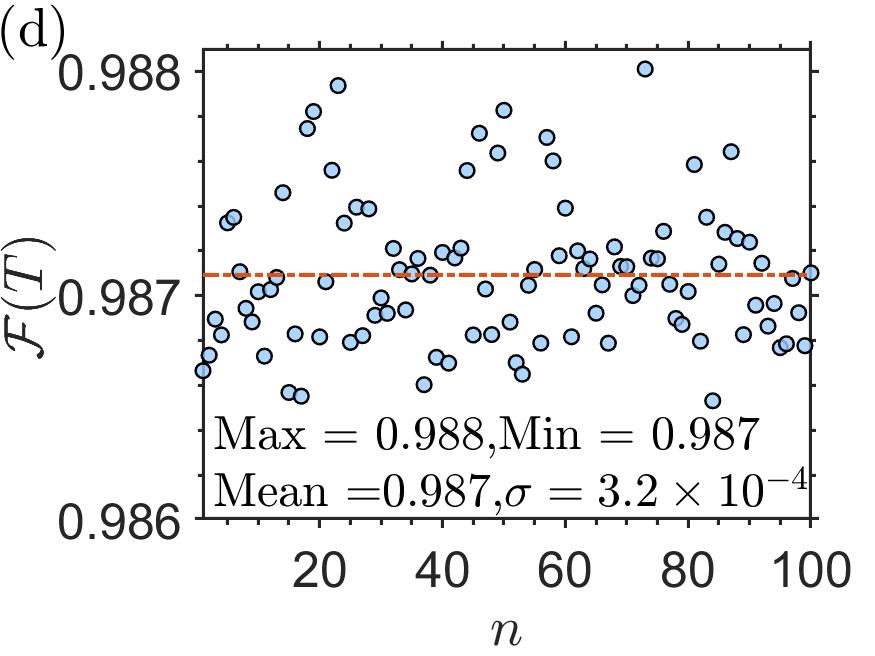}
  \caption{(a) Photonic dynamics during the NAST protocol.
  (b) NAST fidelity versus disorder strength $\delta$ for on-site frequency, inter-resonator hopping and atom-waveguide coupling disorder.
  (c) NAST fidelity as a function of the resonator decay rate $\kappa$ for different atomic energy-relaxation times $T_{1}$.
  (d) NAST fidelities for $100$ randomly sampled input states.
  Parameters are $J=2\pi\times200\,\mathrm{MHz}$, $a=2\pi\times80\,\mathrm{MHz}$, $T=0.1678\,\mathrm{\mu s}$, $N_{c}=41$, $\Delta N=20$, and $N_{a}=3$ in all panels.
  In (d), we set $T_{1}=10\,\mathrm{\mu s}$ and $\kappa=0.5\,\mathrm{MHz}$.}
  \label{non-adiabatic}
\end{figure}

\emph{Robustness}-In the above demonstrations, we have restricted ourselves to the ideal case where disorder and dissipation are neglected.
We show below that the NAST fidelity remains high even under nonideal conditions.

First, we examine the impact of disorder in the on-site resonator frequencies, inter-resonator hopping strengths, and atom--waveguide couplings.
We model each type of disorder by independent Gaussian random variations with zero mean and standard deviation $\delta$.
The NAST fidelity, averaged over 500 disorder realizations, is shown in Fig.~\ref{non-adiabatic}(b).
For all disorder types considered, the fidelity remains above $92\%$ even for a relative fluctuation of $\delta=0.05J\simeq 2\pi\times 10~\mathrm{MHz}$, which lies within currently accessible experimental ranges.
Such robust performance is consistent with the disorder resilience of BICs in waveguide-QED setups~\cite{MW2025}.

Second, we take into account photonic dissipation in the waveguide.
As shown in Fig.~\ref{non-adiabatic}(c), the fidelity remains above $98\%$ even when the resonator decay rate reaches $\kappa=1$~MHz and the atomic energy-relaxation time is on the order of tens of $\mu$s.
The underlying mechanism is captured by Fig.~\ref{imaginary}(a), which illustrates the lifetimes of the eigenstates in the presence of photon loss.
In our NAST protocol, the dynamics predominantly involve only the BICs and their nearest scattering states, which possess the longest lifetimes.
Consequently, high-fidelity transfer can be maintained even in the presence of photonic decay.

Finally, our scheme is applicable to the transfer of an arbitrary state within the single-excitation subspace.
To demonstrate this explicitly, we randomly sample $100$ target states, i.e., random complex amplitudes $C_i$ satisfying $\sum_i |C_i|^2=1$, and plot the corresponding fidelities in Fig.~\ref{non-adiabatic}(d).
We obtain an average fidelity of $\bar{\mathcal{F}}=98.7\%$, with a very narrow distribution of width $\sigma=3.2\times10^{-4}$, which is effectively negligible.
These results confirm that our BIC-based protocol enables high-fidelity ($>98\%$) transfer of \emph{arbitrary, unknown} quantum states.

In the above analyses, we have focused on quantum state transfer for $N_a=3$.
Since the favorable properties of BICs persist for arbitrary $N_a$, as indicated by Eqs.~\eqref{beta} and \eqref{alpha_p}, we expect that our protocol can be straightforwardly extended to larger atomic arrays.
For example, in the SM~\cite{SM} we present results for $N_a=4$, where the NAST fidelity exceeds $97\%$ in the ideal case and remains above $90\%$ when intrinsic dissipation is included.

On the other hand, we also verify in the SM~\cite{SM} that for a system comprising $N_a$ atomic pairs, high-fidelity state transfer cannot be achieved when the parameters are chosen such that the system supports fewer than $N_a$ BICs or no BICs at all.

\emph{Conclusion}-We have demonstrated BICs in a waveguide-QED system where two spatially separated atomic arrays interact with a CRW via time-dependent couplings. When resonator losses are included, these BICs and their nearest scattering states exhibit the smallest imaginary parts of the complex eigenfrequencies, indicating markedly long lifetimes. The BICs are atom--photon dressed states: the photonic component is confined between the two coupling sites in a standing-wave pattern, while the atomic component can encode entanglement. Exploiting these states, we realize the transfer of arbitrary superposition states between the two atomic arrays in both adiabatic and nonadiabatic regimes. The protocol is resilient to multiple forms of disorder, and the fidelity can exceed $98\%$ even in the presence of atomic dissipation and resonator decay.

Most previous state-transfer schemes rely critically on chiral waveguides or nonreciprocal elements. Our BIC-based approach removes this requirement and provides a robust route to high-fidelity state transfer in \emph{bidirectional} waveguide-QED architectures. We anticipate that our results will stimulate further studies of BIC-enabled quantum state engineering and quantum information processing in realistic waveguide platforms with unavoidable imperfections.

This work is supported by Natural Science Foundation of China (Grants Nos. 12375010, 12575011, and  12574397), National Science Fund for Distinguished Young Scholars of China (Grant No. 12425502), the Innovation Program for Quantum Science and Technology (Grant No. 2024ZD0301000) and the National Key Research and Development Program of China (Grant No. 2021YFA1400700).

\clearpage
\newpage
\onecolumngrid


\newcommand\specialsectioning{\setcounter{secnumdepth}{-2}}
\setcounter{equation}{0} \setcounter{figure}{0}

\setcounter{table}{0}
\renewcommand{\theequation}{S\arabic{equation}}
\renewcommand{\thefigure}{S\arabic{figure}}
\renewcommand{\bibnumfmt}[1]{[S#1]}
\renewcommand{\citenumfont}[1]{S#1}
\renewcommand\thesection{S\arabic{section}}
\renewcommand{\baselinestretch}{1.2}

\renewcommand{\theequation}{S\arabic{equation}}


\setcounter{page}{1}\setcounter{secnumdepth}{3} \makeatletter

\begin{center}
	{\Large \textbf{Supplementary Materials for \\ ``Bound state in the continuum and multiple atom state transfer applications in a waveguide QED setup''}}
\end{center}

\maketitle
\onecolumngrid

\vspace{3em}

This Supplementary Material (SM) is organized into three sections. In Sec.~\ref{A}, we provide an analytical proof that the system comprising $N_a$ atomic pairs supports $N_a$ time-independent bound states in the continuum (BICs). In Sec.~\ref{B}, we present the numerical energy spectra, showing that many scattering states lie in the vicinity of each BIC. In Sec.~\ref{C}, we analyze the tunneling dynamics between each BIC and its nearest scattering state during the nonadiabatic state-transfer process. Finally, we show that high-fidelity state transfer cannot be achieved when the system supports fewer than $N_a$ BICs or when BICs are absent.

\section{Bound states in the continuum and their wave functions}$\label{A}$

In this section, we prove that the system supports a set of nondegenerate bound states in the continuum
$\{\ket{\mathrm{BIC}^{(p)}(t)}\}$, which are instantaneous eigenstates of $\hat{H}(t)$ with
\emph{time-independent} eigenvalues $\Omega_p$. We present the conditions required for the existence of these BICs
and derive the analytical form of the wave function $\ket{\mathrm{BIC}^{(p)}(t)}$, which is given as Eq.~(3) in the
main text.

Before analyzing the BICs, we briefly review the system Hamiltonian introduced in Eq.~(2) of the main text.
Under periodic boundary conditions, the CRW is translationally invariant, and the photon annihilation operator in
real space can be expressed as the Fourier transform of the momentum-space operator, reading
$
\hat{a}_{n}=N_{c}^{-1/2}\sum_k e^{-\mathrm{i}kn} \hat{a}_k.\label{Fourier}
$
Here, $N_{c}$ is the number of resonators in the CRW, and $\hat{a}_{k}^{\dagger}$ denotes the bosonic creation
operator with wave vector $k$. We then obtain the Hamiltonian in momentum space
\begin{align}
\hat{H}(t)=&\sum_{k}\omega_{k}\hat{a}_{k}^{\dagger}\hat{a}_{k}
+\sum_{i=1}^{N_{a}}\Omega_{i}
\left(\hat{\sigma}_{+}^{(i)}\hat{\sigma}_{-}^{(i)}+\hat{\sigma}_{+}^{(N_{a}+i)}
\hat{\sigma}_{-}^{(N_{a}+i)}\right)\nonumber \\
&+\sum_{k}\sum_{i=1}^{N_{a}}\Big[\hat{a}_{k}^{\dagger}
\left(\frac{g(t)}{\sqrt{N_{c}}}\hat{\sigma}_{-}^{(i)}e^{ikN_{1}}+
\frac{g'(t)}{\sqrt{N_{c}}}\hat{\sigma}_{-}^{(N_{a}+i)}e^{ikN_{2}}\right)+\text{H.C.}\Big],
\label{Hk(t)}
\end{align}
where $\omega_{k}=\omega_{c}-2J\cos k$ is the dispersion relation of the continuous band supported by the CRW.
In what follows, we set $\omega_{c}=0$ for simplicity.
For a finite CRW with $N_c$ resonators, the allowed wave vectors are
$k\in\{k_{j}=-\pi+(2\pi/N_{c})j\,|\,j=0,1,\dots,N_{c}-1\}$.

Regarding the atomic-frequency configuration, we assume that the $p$th and $(N_a+p)$th atoms share the same
transition frequency $\Omega_p$, which lies inside the continuous band of the CRW, i.e.,
$|\Omega_{p}|<2J$.
The corresponding resonant photon wave vectors are $\pm K_{p}$, determined by
$-2J\cos(\pm K_{p})=\Omega_{p}$ with $K_{p}>0$.
Different atomic pairs are taken to be spectrally distinct, i.e., $\Omega_p\neq\Omega_{q}$ for $p\neq q$.

\subsection{Time-independent eigenfrequency}

In this subsection, we prove that $\hat{H}(t)$ admits a \emph{time-independent} eigenvalue $\Omega_{p}$ when the
separation between the two atomic arrays, $\Delta N=N_{2}-N_{1}$, satisfies $\Delta N K_{p}=m_p\pi$ with
$m_{p}\in\mathbb{Z}^{+}$, and when $\pm K_{p}$ do not belong to the discrete set of photon wave vectors supported by
the finite CRW, i.e., $\pm K_{p}\neq -\pi+(2\pi/N_{c})j$ for all $j=0,1,\dots,N_{c}-1$.
In this case, the dimension of the corresponding eigenspace depends on $N_c$ and $K_p$.

First, we introduce the matrix $\bm{O}(t)$ in the single-excitation basis in momentum space,
\begin{align}
\bm{O}(t)=\hat{H}(t) - \Omega_p \hat{I}=
\begin{bmatrix}
\bm{A} & \bm{B}(t) \\
\bm{B}^{\dagger}(t) & \bm{C}
\end{bmatrix}.
\end{align}
Here,
\begin{equation}
\begin{aligned}
\bm{A} &=
  \operatorname{diag}\bigl(
    \omega_{k_0} - \Omega_p,\,
    \omega_{k_1} - \Omega_p,\,
    \ldots,\,
    \omega_{k_{N_c-1}} - \Omega_p
  \bigr), \\[4pt]
\bm{B} &=
  \bigl[
    \underbrace{\bm{\lambda}_1,\ldots,\bm{\lambda}_1}_{N_a},
    \underbrace{\bm{\lambda}_2,\ldots,\bm{\lambda}_2}_{N_a}
  \bigr],
\quad
\bm{\lambda}_1(t) =
  \frac{g(t)}{\sqrt{N_c}}
  \begin{bmatrix}
    e^{i k_0 N_1} \\[3pt]
    e^{i k_1 N_1} \\[2pt]
    \vdots        \\[2pt]
    e^{i k_{N_c-1} N_1}
  \end{bmatrix},
\quad
\bm{\lambda}_2(t) =
  \frac{g'(t)}{\sqrt{N_c}}
  \begin{bmatrix}
    e^{i k_0 N_2} \\[3pt]
    e^{i k_1 N_2} \\[2pt]
    \vdots        \\[2pt]
    e^{i k_{N_c-1} N_2}
  \end{bmatrix}, \\[4pt]
\bm{C} &=
  \operatorname{diag}\bigl(
    \delta^{(p)}_1,\ldots,\delta^{(p)}_{N_a},
    \delta^{(p)}_1,\ldots,\delta^{(p)}_{N_a}
  \bigr),
\end{aligned}
\end{equation}
where $\delta_{m}^{(p)}=\Omega_{m}-\Omega_{p}$ and $\delta_{m}^{(p)}=0$ if and only if $m=p$.

Next, we introduce the elementary matrix $\bm{M}_{(i,j)}$, whose diagonal elements are all $1$ and whose
off-diagonal elements are zero except for a single $(-1)$ located at the intersection of the $i$th row and
$j$th column. Performing the sequence of elementary transformations
\begin{equation}
\bm{M}=[\prod_{i=1}^{p-1}\bm{M}_{(i,p)}][\prod_{i=p+1}^{N_{a}}\bm{M}_{(i,p)}][\prod_{i=N_{a}+1}^{N_{a}+p-1}\bm{M}_{(i,N_{a}+p)}][\prod_{i=N_{a}+p+1}^{2N_{a}}\bm{M}_{(i,N_{a}+p)}]
\end{equation}
on $\bm{O}(t)$, we obtain a new block matrix $\bm{\tilde{O}}(t)$ with the same rank as $\bm{O}(t)$,
\begin{align}
\bm{\tilde{O}}(t)=\bm{M}\bm{O}(t)\bm{M}^{T}=
\begin{bmatrix}
\bm{A} & \bm{D}(t) \\
\bm{D}^{\dagger}(t) & \bm{C}
\end{bmatrix}.
\end{align}
Here, the $p$th column of $\bm{D}(t)$ is $\bm{\lambda}_{1}$, the $(N_{a}+p)$th column is $\bm{\lambda}_{2}$,
and all other columns are zero. Moreover, $\bm{A}$ is invertible when
$\pm K_{p}\neq -\pi+(2\pi/N_{c})j$ for all $j=0,1,\dots,N_{c}-1$, i.e., when $\pm K_{p}$ do not belong to the
discrete set of photon wave vectors supported by the CRW. In this case, the rank of $\bm{\tilde{O}}(t)$ follows
from the Schur-complement rank theorem~\cite{schur} as
\begin{align}
\text{rank}[\bm{\tilde{O}}(t)] = \text{rank}[\bm{A}] + \text{rank}\left[\bm{C} - \bm{D}^{\dagger}(t)\bm{A}^{-1}\bm{D}(t)\right]
= N_c+\text{rank}\left[\bm{C} - \bm{D}^{\dagger}(t)\bm{A}^{-1}\bm{D}(t)\right].
\end{align}
A direct evaluation of the Schur complement shows that
$\text{rank}\left[\bm{C} - \bm{D}^{\dagger}(t)\bm{A}^{-1}\bm{D}(t)\right]
= 2N_{a}+[\dim(\langle\bm{v}_{1},\bm{v}_{2}\rangle)-2]$, which yields
\begin{align}
\text{rank}[\hat{H}(t) - \Omega_p \hat{I}]
= N_{c}+2N_{a}+\big[\dim(\langle\bm{v}_{1},\bm{v}_{2}\rangle)-2\big],
\end{align}
where
\begin{align}
\bm{v}_{1}=\begin{bmatrix}
\frac{g(t)^{2}}{N_{c}}\sum_{k}\frac{1}{\Omega_{p}-\omega_{k}}\\
\frac{g(t)g'(t)}{N_{c}}\sum_{k}\frac{e^{-ik\Delta N}}{\Omega_{p}-\omega_{k}}
\end{bmatrix},
\qquad
\bm{v}_{2}=\begin{bmatrix}
\frac{g(t)g'(t)}{N_{c}}\sum_{k}\frac{e^{ik\Delta N}}{\Omega_{p}-\omega_{k}}\\
\frac{g'(t)^{2}}{N_{c}}\sum_{k}\frac{1}{\Omega_{p}-\omega_{k}}
\end{bmatrix},
\end{align}
and $\langle\bm{v}_{1},\bm{v}_{2}\rangle$ denotes the space spanned by $\bm{v}_{1}$ and $\bm{v}_{2}$.

Since the interference condition $K_p\Delta N = m_p\pi~(m_{p}\in\mathbb{Z}^{+})$ is satisfied, we have
\begin{align}
&\frac{1}{N_{c}}\sum_{k}\frac{1}{\Omega_{p}-\omega_{k}}=
\begin{cases}
-\frac{1}{2\sin K_{p}}\tan(\frac{N_{c}K_{p}}{2}),\ N_{c}\equiv1 \pmod{2},\\
\frac{1}{2\sin K_{p}}\cot(\frac{N_{c}K_{p}}{2}),\ N_{c}\equiv0 \pmod{2},
\end{cases}\\
&\frac{1}{N_{c}}\sum_{k}\frac{e^{\pm ik\Delta N}}{\Omega_{p}-\omega_{k}}=
\begin{cases}
-\frac{(-1)^{m_{p}}}{2\sin K_{p}}\tan(\frac{N_{c}K_{p}}{2}),\ N_{c}\equiv1 \pmod{2},\\
\frac{(-1)^{m_{p}}}{2\sin K_{p}}\cot(\frac{N_{c}K_{p}}{2}),\ N_{c}\equiv0 \pmod{2}.
\end{cases}
\end{align}
This implies
\begin{align}
\dim(\langle\bm{v}_{1},\bm{v}_{2}\rangle) =
\begin{cases}
0, &
\begin{aligned}
&N_{c}K_{p}=2n\pi,\ N_{c}\equiv1 \pmod{2}, \\
&\text{or } N_{c}K_{p}=(2n-1)\pi,\ N_{c}\equiv0 \pmod{2},
\end{aligned} \\[4pt]
1, &
\begin{aligned}
&N_{c}K_{p}\neq2n\pi,\ N_{c}\equiv1 \pmod{2}, \\
&\text{or } N_{c}K_{p}\neq(2n-1)\pi,\ N_{c}\equiv0 \pmod{2},
\end{aligned}
\end{cases}
\label{condition}
\end{align}
with $n\in\mathbb{Z}^{+}$. When $\dim(\langle\bm{v}_{1},\bm{v}_{2}\rangle)=0$, the eigenspace associated with
$E(t)=\Omega_{p}$ is two-dimensional, yielding two degenerate eigenstates at $\Omega_p$. By contrast, when
$\dim(\langle\bm{v}_{1},\bm{v}_{2}\rangle)=1$, the eigenspace is one-dimensional, indicating a unique eigenstate
with the time-independent eigenvalue $E(t)=\Omega_p$.

\subsection{BIC wave function}

In this subsection, we focus on the second scenario in Eq.~(\ref{condition}). We show that the nondegenerate
eigenstate with a time-independent eigenfrequency $\Omega_p$ indeed corresponds to a BIC by explicitly deriving
the analytical form of its wave function. Note that the condition
$\dim(\langle\bm{v}_{1},\bm{v}_{2}\rangle)=1$ implies that $\bm{v}_1$ and $\bm{v}_2$ are linearly dependent, i.e.,
\begin{align}
(-1)^{m_{p}+1}\frac{g'(t)}{g(t)}\,\bm{v}_{1}+\bm{v}_{2}=0,\qquad \bm{v}_{1,2}\neq\bm{0}.
\label{line_dependent}
\end{align}

We begin by expanding the corresponding instantaneous eigenstate as
\begin{equation}
\ket{\psi^{(p)}(t)}=[\sum_{k}\alpha_{k}^{(p)}(t)\hat{a}_{k}^{\dagger}+\sum_{i=1}^{N_{a}}(\beta_{i}^{(p)}(t)\hat{\sigma}_{+}^{(i)}+\beta_{N_{a}+i}^{(p)}(t)\hat{\sigma}_{+}^{(N_{a}+i)})]\ket{G},
\end{equation}
where $\ket{G}$ denotes the ground state of the combined CRW--atom system. Using the instantaneous eigenequation
$\hat{H}(t)\ket{\psi^{(p)}(t)}=\Omega_{p}\ket{\psi^{(p)}(t)}$, we obtain
\begin{align}
&(\Omega_{p}-\omega_{k})\alpha_{k}^{(p)}(t)=\sum_{i=1}^{N_{a}}(\frac{g(t)}{\sqrt{N_{c}}}e^{ikN_{1}}\beta_{i}^{(p)}(t)+\frac{g'(t)}{\sqrt{N_{c}}}e^{ikN_{2}}\beta_{N_{a}+i}^{(p)}(t)),\label{E_alphak}\\
&(\Omega_{p}-\Omega_{i})\beta_{i}^{(p)}(t)=\sum_{k}\frac{g(t)}{\sqrt{N_{c}}}e^{-ikN_{1}}\alpha_{k}^{(p)}(t),\qquad(i=1,2,\dots,N_{a}),\label{E_betai}\\
&(\Omega_{p}-\Omega_{i})\beta_{N_{a}+i}^{(p)}(t)=\sum_{k}\frac{g'(t)}{\sqrt{N_{c}}}e^{-ikN_{2}}\alpha_{k}^{(p)}(t).\qquad(i=1,2,\dots,N_{a})\label{E_betaNa+i}
\end{align}
To simplify notation, we omit the explicit time dependence of the variables below. From Eq.~(\ref{E_alphak}), we have
\begin{equation}
\alpha_{k}^{(p)}=\frac{1}{\sqrt{N_{c}}}\sum_{i=1}^{N_{a}}\frac{ge^{ikN_{1}}\beta_{i}^{(p)}+g'e^{ikN_{2}}\beta_{N_{a}+i}^{(p)}}{\Omega_{p}-\omega_{k}}.\label{alphak}
\end{equation}
Substituting Eq.~(\ref{alphak}) into Eqs.~(\ref{E_betai}) and (\ref{E_betaNa+i}), we obtain
\begin{align}
(\Omega_{p}-\Omega_{i})\beta_{i}^{(p)}=\sum_{j=1}^{N_{a}}\Big[\big(\frac{1}{N_{c}}\sum_{k}\frac{1}{\Omega_{p}-\omega_{k}}\big)g^{2}\beta_{j}^{(p)}+\big(\frac{1}{N_{c}}\sum_{k}\frac{e^{ik\Delta N}}{\Omega_{p}-\omega_{k}}\big)gg'\beta_{N_{a}+j}^{(p)}\Big]\equiv C_{1},\label{C1}\\
(\Omega_{p}-\Omega_{i})\beta_{N_{a}+i}^{(p)}=\sum_{j=1}^{N_{a}}\Big[\big(\frac{1}{N_{c}}\sum_{k}\frac{e^{-ik\Delta N}}{\Omega_{p}-\omega_{k}}\big)gg'\beta_{j}^{(p)}+\big(\frac{1}{N_{c}}\sum_{k}\frac{1}{\Omega_{p}-\omega_{k}}\big)g'^{2}\beta_{N_{a}+j}^{(p)}\Big]\equiv C_{2}.
\end{align}
The right-hand sides of the above two equations are independent of $i$. Setting $i=p$, we obtain
$C_1=C_2\equiv 0$. Therefore, $(\Omega_{p}-\Omega_{i})\beta_{i}^{(p)}=(\Omega_{p}-\Omega_{i})\beta_{N_{a}+i}^{(p)}\equiv 0$
holds for all $i=1,2,\dots,N_{a}$. For $i\neq p$, we have $\Omega_{i}\neq\Omega_{p}$, which implies
\begin{equation}
\beta_{i}^{(p)}\equiv0\ (i\neq p,N_{a}+p).
\label{beta_neq0}
\end{equation}
Substituting Eq.~(\ref{beta_neq0}) into Eq.~(\ref{C1}), we obtain
\begin{align}
\frac{g(t)^{2}}{N_{c}}\sum_{k}\frac{1}{\Omega_{p}-\omega_{k}}\beta_{p}^{(p)}+\frac{g(t)g'(t)}{N_{c}}\sum_{k}\frac{e^{ik\Delta N}}{\Omega_{p}-\omega_{k}}\beta_{N_{a}+p}^{(p)}=0.
\end{align}

Using the linear-dependence condition in Eq.~(\ref{line_dependent}), we further find
\begin{align}
\frac{\beta_{p}^{(p)}}{\beta_{N_{a}+p}^{(p)}}=(-1)^{m_{p}+1}\frac{g'(t)}{g(t)}
\ \Rightarrow\
\beta_{N_{a}+p}^{(p)} = (-1)^{m_{p}+1} \frac{g(t)}{g'(t)} \beta_{p}^{(p)}.
\label{beta_ratio}
\end{align}
With this relation, $\ket{\psi^{(p)}(t)}$ can be rewritten in the real-space basis as
\begin{align}
\ket{\psi^{(p)}(t)}=[\sum_{n}\alpha_{n}^{(p)}(t)\hat{a}_{n}^{\dagger}
+\beta_{p}^{(p)}(t)\hat{\sigma}_{+}^{(p)}
+(-1)^{m_{p}+1}\frac{g}{g'}\beta_{p}^{(p)}(t)\hat{\sigma}_{+}^{(N_{a}+p)}]\ket{G},
\end{align}
where $\alpha_{n}^{(p)}$ is obtained from the Fourier transform of $\alpha_{k}^{(p)}$, with
$\alpha_{n}^{(p)}=(\sqrt{N_{c}})^{-1/2}\sum_{k}\alpha_{k}^{(p)}e^{-ikn}$.
Substituting Eqs.~(\ref{alphak}), (\ref{beta_neq0}), and (\ref{beta_ratio}) yields
\begin{align}
\alpha_{n}^{(p)}=g\Big[\dfrac{1}{N_{c}}\sum_{k}\frac{e^{ik(N_{1}-n)}+(-1)^{m_{p}+1}e^{ik(N_{2}-n)}}{\Omega_{p}-\omega_{k}}\Big]\beta_{p}^{(p)}=
\begin{cases}
0, & n<N_{1},\\
\dfrac{\sin{[(n-N_{1})K_{p}]}}{\sin{K_{p}}}\dfrac{g}{J}\beta_{p}^{(p)}, & N_{1}\leq n\leq N_{2},\\
0, & n>N_{2},
\end{cases}
\label{photon_distribution}
\end{align}
and, after normalization,
\begin{align}
\beta_{p}^{(p)}(t)=\frac{1}{\sqrt{1+[g(t)/g'(t)]^2+\Delta N [g(t)/(\sqrt{2}J\sin{K_{p}})]^2}}.
\end{align}

These results yield the instantaneous eigenstate at $E(t)=\Omega_p$ as
\begin{align}
\ket{\psi^{(p)}(t)}=\beta_{p}^{(p)}(t)\Big[\sum_{n=N_{1}}^{N_{2}}\frac{\sin{[(n-N_{1})K_{p}]}}{{\sin{K_{p}}}}\frac{g(t)}{J}\hat{a}_{n}^{\dagger}
+\hat{\sigma}_{+}^{(p)}+(-1)^{m_{p}+1}\frac{g(t)}{g^{\prime}(t)}\hat{\sigma}_{+}^{(N_{a}+p)}\Big]\ket{G},
\label{expansion}
\end{align}
which is equivalent to Eqs.~(3)--(5) in the main text. The photonic component is confined to the region
$N_1\le n\le N_2$ [Eq.~(\ref{photon_distribution})], demonstrating real-space localization. Since the
eigenvalue $\Omega_p$ lies within the continuous band, we identify
$\ket{\psi^{(p)}(t)}\equiv\ket{{\rm BIC}^{(p)}(t)}$ as a bound state in the continuum.

\section{Numerical eigenfrequency spectrum}\label{B}

\begin{figure}
  \centering
  \includegraphics[width=0.48\columnwidth]{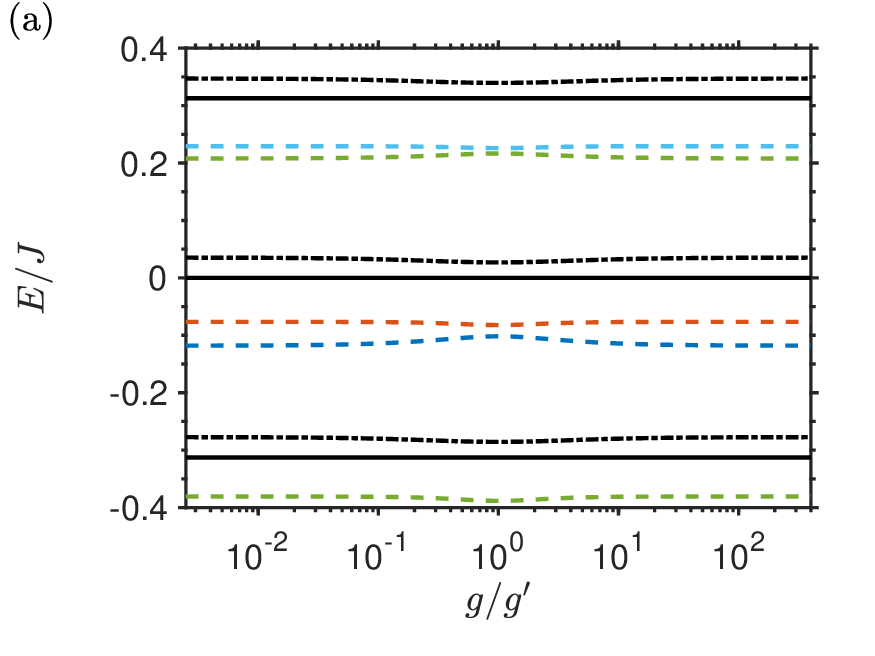}\includegraphics[width=0.48\columnwidth]{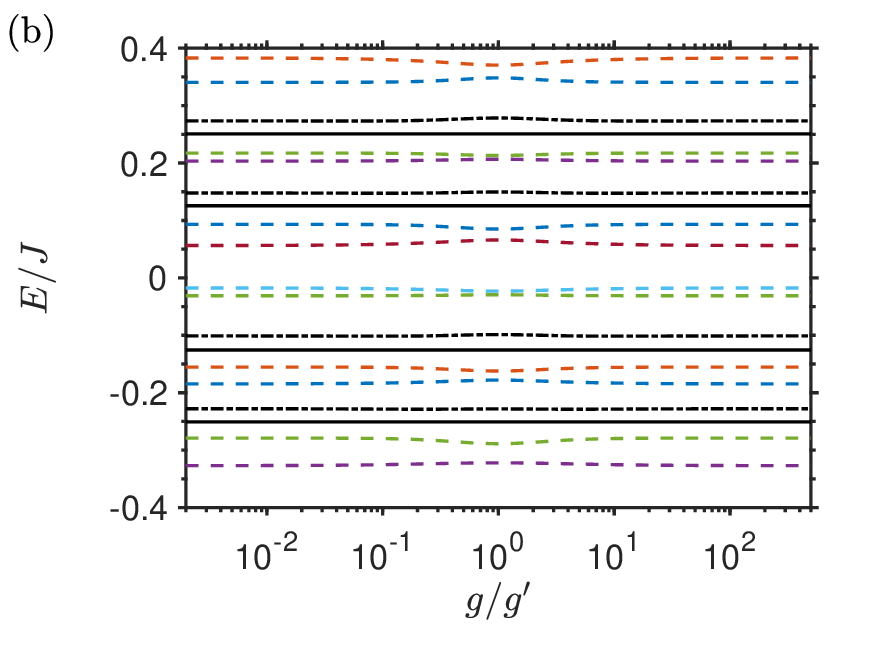}
  \caption{(a) and (b) Partial energy spectra for $N_{a}=3$ and $N_{a}=4$, respectively. Solid lines represent BICs, dash-dotted lines denote the nearest scattering states, and dashed lines indicate other scattering states.}
  \label{spectrum}
\end{figure}

We have analytically shown that, for all $p=1,2,\ldots,N_{a}$, the system supports $N_{a}$ BICs whose
time-independent, nondegenerate eigenvalues coincide with the corresponding atomic-pair transition
frequencies when the following conditions are satisfied:
$\pm K_{p}\notin \{k_{j}=-\pi+(2\pi/N_{c})j \mid j=0,1,\ldots,N_{c}-1\}$,
$K_{p}\Delta N=m_{p}\pi$, and either
$[N_{c}\ \text{odd and}\ N_{c}K_{p}\neq 2n\pi]$ or
$[N_{c}\ \text{even and}\ N_{c}K_{p}\neq (2n-1)\pi]$ (with $m_p,n\in\mathbb{Z}^{+}$).
To corroborate these results, we numerically plot the spectra of each BIC and several nearby scattering states
for $N_a=3$ and $N_a=4$ in Fig.~\ref{spectrum}(a) and (b), respectively. In both cases, the scattering-state
frequencies vary with time as the couplings are modulated, while the BIC eigenfrequencies remain fixed.
Moreover, a sizable gap is clearly visible between each BIC and its nearest scattering state.

These gaps enable adiabatic state transfer (AST) between the two atomic arrays. In this process, the couplings
$g(t)$ and $g'(t)$ must be varied sufficiently slowly to suppress tunneling between the BIC manifold and the
scattering spectrum. Although AST can in principle achieve perfect state transfer with unit fidelity, the
required time can become comparable to the coherence time of realistic platforms such as transmon
qubits~\cite{transmon}. For example, the main text shows in Fig.~2(b) that the required time for $N_a=3$ is
approximately $1.6\,\mu\mathrm{s}$. Because the gap between the BIC and its nearest scattering state becomes
smaller for $N_a=4$, the corresponding AST time increases to $9.9734\,\mu\mathrm{s}$. To better mitigate
decoherence, we therefore turn to the nonadiabatic state-transfer (NAST) protocol.

\section{Tunneling dynamics in non-adiabatic state transfer process}\label{C}

To further shorten the transfer time beyond the AST process, we resort to the NAST protocol, which permits
controlled tunneling between the BICs and their nearest scattering states. To quantitatively describe the
tunneling dynamics during state transfer, we express the time-dependent state as
\begin{align}
\ket{\phi(t)}=\sum_{m=1}^{N_{a}} C_{m} \left[ \tilde{C}_{B}^{(m)}(t) \ket{\text{BIC}^{(m)}(t)} +
\tilde{C}_{S}^{(m)}(t) \ket{\text{SC}^{(m)}(t)} \right],
\end{align}
with initial conditions $\tilde{C}_{B}^{(m)}(0) = 1$ and $\tilde{C}_{S}^{(m)}(0) = 0$ for all
$m = 1,2,\dots,N_{a}$. From the Schr\"odinger equation
$i\partial_t \ket{\phi(t)} = \hat{H}(t) \ket{\phi(t)}$, we obtain
\begin{align}
&i\sum_{m} C_{m} \left[ \dot{\tilde{C}}_{B}^{(m)}(t) \ket{\text{BIC}^{(m)}(t)}
+ \tilde{C}_{B}^{(m)}(t) \partial_t \ket{\text{BIC}^{(m)}(t)} \right. \nonumber \\
&\quad \left. + \dot{\tilde{C}}_{S}^{(m)}(t) \ket{\text{SC}^{(m)}(t)}
+ \tilde{C}_{S}^{(m)}(t) \partial_t \ket{\text{SC}^{(m)}(t)} \right] \nonumber \\
&= \sum_{m} C_{m} \left[ \tilde{C}_{B}^{(m)}(t) \Omega_{m} \ket{\text{BIC}^{(m)}(t)}
+ \tilde{C}_{S}^{(m)}(t) E_{S}^{(m)}(t) \ket{\text{SC}^{(m)}(t)} \right],
\end{align}
where $E_{S}^{(m)}(t)$ denotes the instantaneous eigenvalue of $\ket{\text{SC}^{(m)}(t)}$. Projecting onto
$\bra{\text{BIC}^{(p)}(t)}$ and $\bra{\text{SC}^{(p)}(t)}$, we arrive at the closed coupled equations
\begin{align}
\dot{\tilde{C}}_{B}^{(p)}(t) + i\Omega_{p} \tilde{C}_{B}^{(p)}(t) + F_p(t)\tilde{C}_{S}^{(p)}(t) &= 0, \nonumber \\
\dot{\tilde{C}}_{S}^{(p)}(t) + iE_{S}^{(p)}(t) \tilde{C}_{S}^{(p)}(t) - F_p(t)\tilde{C}_{B}^{(p)}(t) &= 0 .
\end{align}

\begin{figure}
  \centering
  \includegraphics[width=0.3\columnwidth]{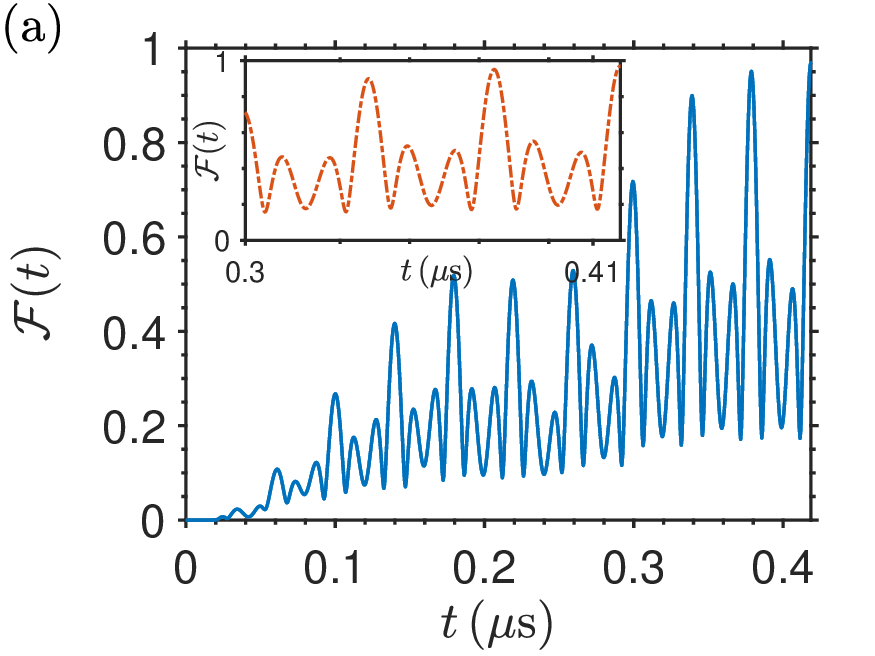}
  \includegraphics[width=0.3\columnwidth]{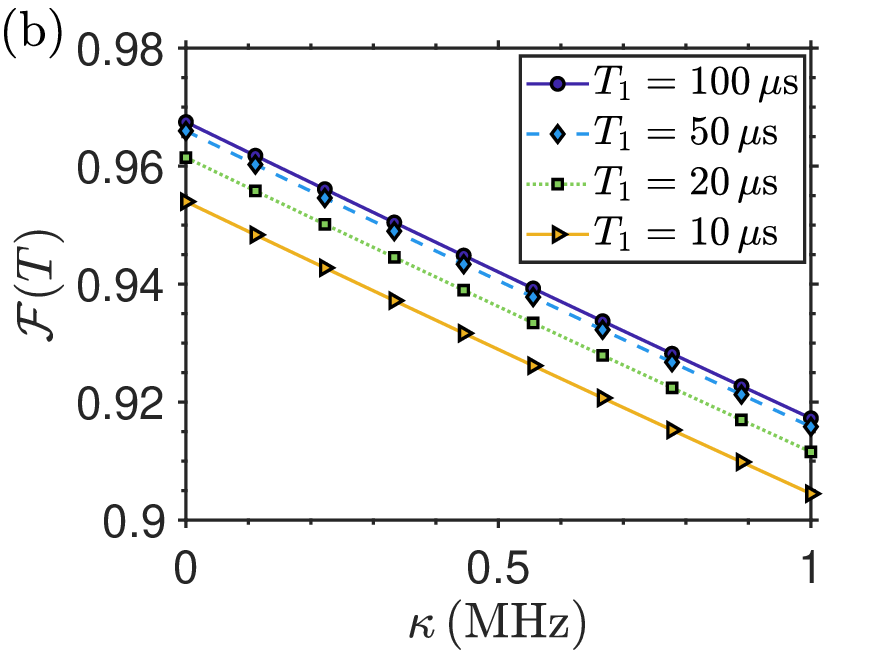}
  \includegraphics[width=0.3\columnwidth]{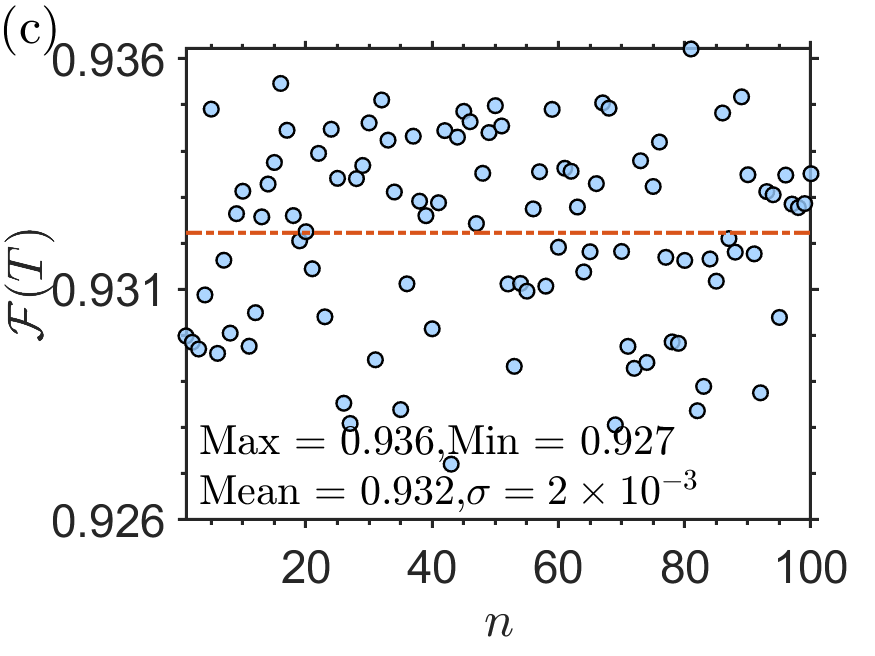}
  \caption{(a) Dynamics of the NAST fidelity. (b) NAST fidelity as a function of the resonator decay rate $\kappa$
  for different atomic energy-relaxation times $T_{1}$. (c) NAST fidelities for $100$ arbitrary input states.
  The parameters are $J=2\pi\times200\,{\rm MHz}$, $a=0.5J$, $T=0.4187\,{\rm \mu s}$, $N_{c}=101$, $\Delta N=50$,
  and $N_{a}=4$ in all panels. The state is chosen as
  $C_{1} = \sqrt{5}/5$, $C_{2} = (\sqrt{10}/5)e^{i\pi/5}$, $C_{3} = (\sqrt{5}/5)e^{i\pi/3}$,
  and $C_{4} = (\sqrt{5}/5)e^{i\pi/10}$ in (a) and (b). In (c), we set $T_{1}=10\,{\rm \mu s}$ and
  $\kappa=0.5\,{\rm MHz}$.}
  \label{Na4_non-adiabatic}
\end{figure}

Here, $F_p(t) = \langle \text{BIC}^{(p)}(t) | \partial_t | \text{SC}^{(p)}(t) \rangle$ characterizes the
effective coupling between these two instantaneous eigenstates induced by the time dependence of $\hat{H}(t)$.
To obtain the above equations, we have assumed that
$\langle \text{BIC}^{(p)}(t) | \partial_t | \text{SC}^{(m)}(t) \rangle = 0$ for $m \neq p$.
Physically, this assumption corresponds to an intermediate driving-timescale regime: the modulation is fast
enough to allow tunneling between each BIC and its nearest scattering state, yet not so fast as to activate
significant transitions to non-nearest scattering states.

The solution of the coupled equations is uniquely determined by the coefficients and initial conditions, while
$\Omega_{p}$, $E_{S}^{(p)}(t)$, and $F_p(t)$ are fixed by the intrinsic properties of $\hat{H}(t)$. Crucially,
this implies that the evolution of $\tilde{C}_{B,S}^{(p)}(t)$ is independent of the choice of the
superposition coefficients $\{C_{p}\}$ that specify the initial atomic state. Consequently, as long as a time
$T$ exists such that
\begin{align}
(-1)^{m_{p}+1} \tilde{C}_{B}^{(p)}(T) \approx e^{i\Phi}, \qquad \tilde{C}_{S}^{(p)}(T) \approx 0,
\end{align}
for all $p=1,2,\dots,N_{a}$, we can complete the transfer of an arbitrary atomic state up to a global phase.

In the main text, we have shown that for $N_a=3$ the required time is shortened from
$1.6\,{\rm \mu s}$ in AST to $0.16\,{\rm \mu s}$ in NAST. Here, we further demonstrate that the NAST protocol
also performs well for $N_a=4$, with a required time of $T\simeq 0.41\,{\rm \mu s}$, as shown in
Fig.~\ref{Na4_non-adiabatic}(a). For this larger system, the NAST protocol remains robust against atomic
dissipation and photon loss. As shown in Fig.~\ref{Na4_non-adiabatic}(b), for the fixed input state given in the
caption, the fidelity can be kept above $90\%$ over a broad range of $\kappa$ and $T_1$. Moreover, for $100$
randomly sampled states, the average fidelity reaches $93.2\%$ with a negligible distribution width under
moderate photon decay $\kappa=0.5\,{\rm MHz}$ and atomic decoherence time $T_1=10\,{\rm \mu s}$, as shown in
Fig.~\ref{Na4_non-adiabatic}(c).

\section{Failure transfer with fewer or without BIC}\label{D}

In the main text and the previous sections of this SM, we have demonstrated that for a system consisting of
$N_a$ pairs of atoms, nearly perfect transfer of an arbitrary atomic state can be achieved provided that the
system supports $N_a$ BICs. Here we show that this is no longer the case when fewer than $N_a$ BICs exist, or
when all BICs are absent. To this end, we change only the separation between the two atomic arrays $\Delta N$,
while keeping all other parameters unchanged.

\begin{figure}
  \centering
  \includegraphics[width=0.48\columnwidth]{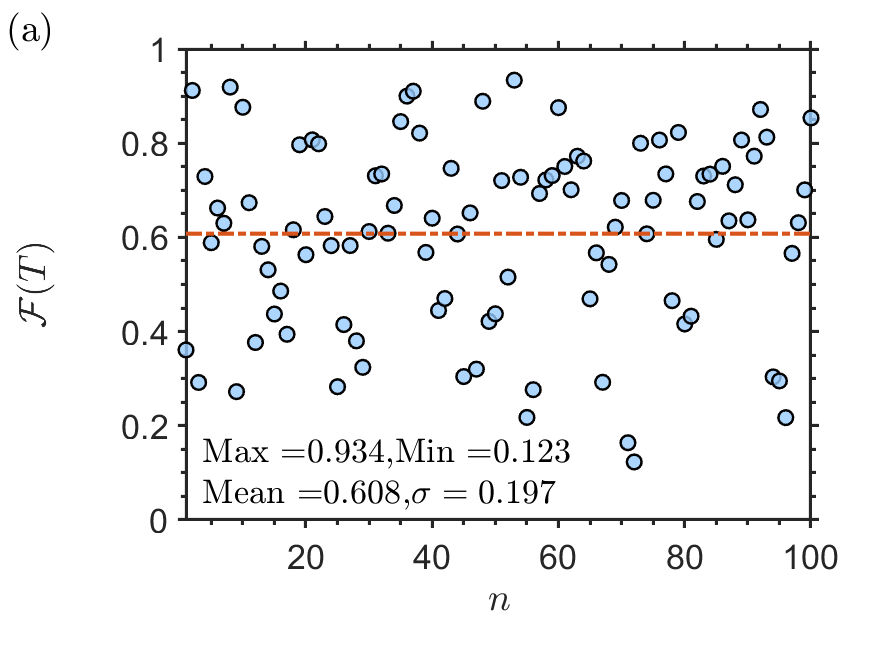}\includegraphics[width=0.48\columnwidth]{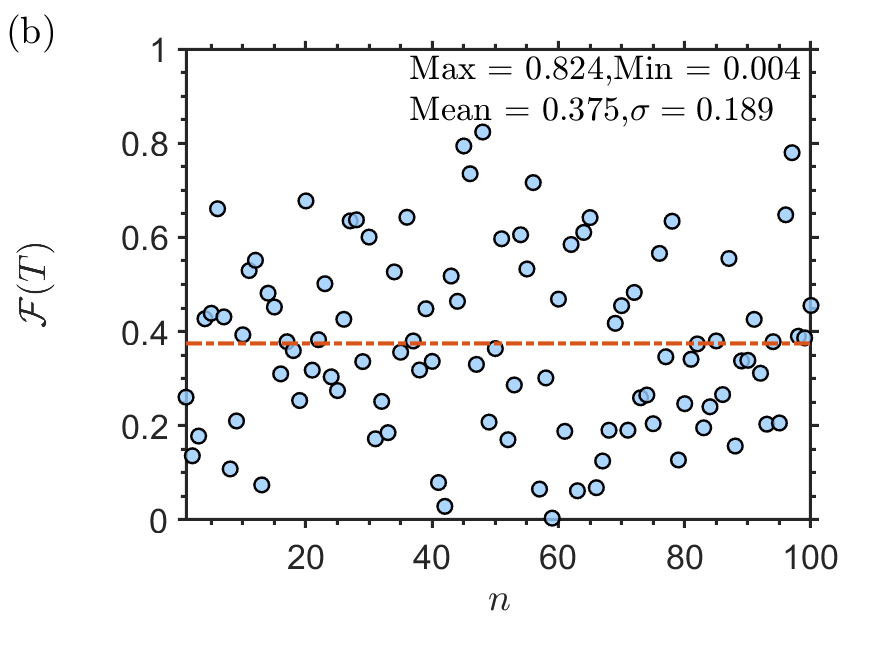}
  \caption{(a) NAST fidelities for 100 arbitrary input states when $N_a=3$ with only one remaining BIC.
  (b) NAST fidelities for 100 arbitrary input states when $N_a=4$ without BICs.
  Parameters are $J=2\pi\times200\,\text{MHz}$, $a=0.4J$, $T=0.1678\,\mu\text{s}$, $N_c=41$, and $\Delta N=26$
  for $N_a=3$ in (a); and $J=2\pi\times200\,\text{MHz}$, $a=0.5J$, $T=0.4187\,\mu\text{s}$, $N_c=101$, and
  $\Delta N=8$ for $N_a=4$ in (b). We set $T_1=10\,\mu\text{s}$ and $\kappa=0.5\,\text{MHz}$ in all panels.}
  \label{nonBIC}
\end{figure}

For $N_a=3$, we fix $N_c=41$ and set $\Delta N=26$. In this case, the system supports only one BIC at
$E_{\rm BIC}=0$, while the other two BICs discussed in the main text merge into the scattering continuum.
As shown in Fig.~\ref{nonBIC}(a), for $100$ randomly chosen input states, the mean fidelity is only
$\bar{\mathcal{F}}=60.8\%$, with a broad distribution width $\sigma=0.197$.
For $N_a=4$, all four BICs become scattering states when we change $\Delta N=50$ to $\Delta N=8$.
Consequently, the mean fidelity further decreases to $\bar{\mathcal{F}}=37.5\%$, as shown in
Fig.~\ref{nonBIC}(b). In these cases, high-fidelity state transfer between the two atomic arrays cannot be
realized.

\begin{thebibliography}{99}
\bibitem{DR2017}D. Roy, C. M. Wilson, and O. Firstenberg, Colloquium: Strongly interacting photons in one-dimensional continuum, Rev. Mod. Phys. \textbf{89}, 021001 (2017).
\bibitem{DE2018}D. E. Chang, J. S. Douglas, A. Gonzalez-Tudela, C.-L. Hung, and H. J. Kimble, Colloquium: Quantum matter built from nanoscopic lattices of atoms and photons, Rev. Mod. Phys. \textbf{90}, 031002 (2018).
\bibitem{AS2023_1}A. S. Sheremet, M. I. Petrov, I. V. Iorsh, A. V. Poshakinskiy, and A. N. Poddubny, Waveguide quantum electrodynamics: Collective radiance and photon-photon correlations, Rev. Mod. Phys. \textbf{95}, 015002 (2023).
\bibitem{XG2017}X. Gu, A. F. Kockum, A. Miranowicz, Y.-X. Liu, and F. Nori, Microwave photonics with superconducting quantum circuits, Phys. Rep. \textbf{718}, 1 (2017).
\bibitem{IC2020}I. Carusotto, A. A. Houck, A. J. Koll\'{a}r, P. Roushan, D. I. Schuster, and J. Simon, Photonic materials in circuit quantum electrodynamics, Nat. Phys. \textbf{16}, 268 (2020).
\bibitem{NG2020}N. Gheeraert, S. Kono, and Y. Nakamura, Programmable directional emitter and receiver of itinerant microwave photons in a waveguide, Phys. Rev. A \textbf{102}, 053720 (2020).
\bibitem{AB2021}A. Blais, A. L. Grimsmo, S. M. Girvin, and A. Wallraff, Circuit quantum electrodynamics, Rev. Mod. Phys. \textbf{93}, 025005 (2021).
\bibitem{JC2008}J. Clarke, and F. K. Wilhelm, Superconducting quantum bits, Nature \textbf{453}, 1031 (2008).
\bibitem{GW2017}G Wendin, Quantum information processing with superconducting circuits: a review, Rep. Prog. Phys. \textbf{80}, 106001 (2017).
\bibitem{AB2020}A. Blais, S. M. Girvin, and  W. D. Oliver, Quantum information processing and quantum optics with circuit quantum electrodynamics, Nat. Phys. \textbf{16}, 247 (2020).
\bibitem{MK2020}M. Kjaergaard, M. E. Schwartz, J. Braum\"{u}ller, P. Krantz, J. I.-J. Wang, S. Gustavsson, and W. D. Oliver, Superconducting Qubits: Current State of Play, Annu. Rev. Condens. Matter Phys. \textbf{11}, 369 (2020).
\bibitem{TS2009}T. Shi, and C. P. Sun, Lehmann-Symanzik-Zimmermann reduction approach to multiphoton scattering in coupledresonator arrays, Phys. Rev. B \textbf{79}, 205111 (2009).
\bibitem{GC2016}G. Calaj\'{o}, F. Ciccarello, D. Chang, and P. Rabl, Atom-field dressed states in slow-light waveguide QED, Phys. Rev. A \textbf{93}, 033833 (2016).
\bibitem{YL2017}Y. Liu and A. A. Houck, Quantum electrodynamics near a photonic band gap, Nat. Phys. \textbf{13}, 48 (2017).
\bibitem{LK2018}L. Krinner, M. Stewart, A. Pazmi\~{n}o, J. Kwon, and D. Schneble, Spontaneous emission of matter waves from a tunable open quantum system, Nature (London) \textbf{559}, 589 (2018).
\bibitem{WZ2020}W. Zhao and Z. Wang, Single-photon scattering and bound states in an atom-waveguide system with two or multiple coupling points, Phys. Rev. A \textbf{101}, 053855 (2020).
\bibitem{AS2023_2}A. Soro, C. S. Mu\~{n}oz, and A. F. Kockum, Interaction between giant atoms in a one-dimensional structured environment, Phys. Rev. A \textbf{107}, 013710 (2023).
\bibitem{CJ2016}C.-J. Yang, and J.-H. An, Resonance fluorescence beyond the dipole approximation of a quantum dot in a plasmonic nanostructure, Phys. Rev. A \textbf{93}, 053803 (2016).
\bibitem{QJ2010}Q.-J. Tong, J.-H. An, H.-G. Luo, and C. H. Oh, Mechanism of entanglement preservation, Phys. Rev. A \textbf{81}, 052330 (2010).
\bibitem{AW2012}A. W. Chin, S. F. Huelga, and M. B. Plenio, Quantum Metrology in Non-Markovian Environments, Phys. Rev. Lett. \textbf{109}, 233601 (2012).
\bibitem{KB2019}K. Bai, Z. Peng, H.-G. Luo and J.-H. An, Retrieving ideal precision in noisy quantum optical metrology, Phys. Rev. Lett. \textbf{123}, 040402 (2019).
\bibitem{FH2019}F. H. Stillinger, and D. R. Herrick, Bound states in the continuum, Phys. Rev. A \textbf{11}, 446 (1975).
\bibitem{DC2008}D. C. Marinica, A. G. Borisov, and S. V. Shabano, Bound states in the continuum in photonics, Phys. Rev. Lett. \textbf{100}, 183902 (2008).
\bibitem{MI2012}M. I. Molina, A. E. Miroshnichenko, and Y. S. Kivshar, Surface bound states in the continuum, Phys. Rev. Lett. \textbf{108}, 070401 (2012).
\bibitem{SI2021}S. I. Azzam, and A. V. Kildishev, Photonic bound states in the continuum: from basics to applications, Adv. Opt. Mater. \textbf{9}, 2001469 (2021).
\bibitem{QQ2023}Q.-Y. Qiu, Y. Wu, and X.-Y. L\"{u}, Collective radiance of giant atoms in non-Markovian regime Sci. China Phys. Mech. Astron. \textbf{66}, 224212 (2023).
\bibitem{MK2023}M. Kang, T. Liu, C. T. Chan, and M. Xiao, Applications of bound states in the continuum in photonics, Nat. Rev. Phys. \textbf{5}, 659 (2023).
\bibitem{XZ2023w}X. Zhang, C. Liu, Z. Gong, and Z. Wang, Quantum interference and controllable magic cavity QED via a giant atom in a coupled resonator waveguide, Phys. Rev. A \textbf{108}, 013704 (2023).
\bibitem{JI1997}J. I. Cirac, P. Zoller, H. J. Kimble, and H. Mabuchi, Quantum State Transfer and Entanglement Distribution among Distant Nodes in a Quantum Network, Phys. Rev. Lett. \textbf{78}, 3221 (1997).
\bibitem{KI2016}K. Inomata, Z. Lin, K. Koshino, W. D. Oliver, J.-S. Tsai, T. Yamamoto, and Y. Nakamura, Single microwave-photon detector using an artificial $\Lambda$-type three-level system, Nat. Commun. \textbf{7}, 12303 (2016).
\bibitem{ZL2017}Z.-L. Xiang, M. Zhang, L. Jiang, and P. Rabl, Intracity Quantum Communication via Thermal Microwave Networks, Phys. Rev. X \textbf{7}, 011035 (2017).
\bibitem{PK2018}P. Kurpiers, P. Magnard, T. Walter, B. Royer, M. Pechal, J. Heinsoo, Y. Salath\'{e}, A. Akin, S. Storz, J.-C. Besse, S. Gasparinetti, A. Blais, and A. Wallraff, Deterministic quantum state transfer and remote entanglement using microwave photons, Nature \textbf{558}, 264 (2018).
\bibitem{CJ2018}C. J. Axline, L. D. Burkhart, W. Pfaff, M. Zhang, K. Chou, P. Campagne-Ibarcq, P. Reinhold, L. Frunzio, S. M. Girvin, L. Jiang, M. H. Devoret, and R. J. Schoelkopf, On-demand quantum state transfer and entanglement between remote microwave cavity memories, Nature Phys. \textbf{14}, 705 (2018).
\bibitem{PM2020}P. Magnard, S. Storz, P. Kurpiers, J. Sch\"{a}r, F. Marxer, J. L\"{u}tolf, T. Walter, J.-C. Besse, M. Gabureac, K. Reuer, A. Akin, B. Royer, A. Blais, and A. Wallraff, Microwave Quantum Link between Superconducting Circuits Housed in Spatially Separated Cryogenic Systems, Phys. Rev. Lett. \textbf{125}, 260502 (2020).
\bibitem{BV2017}B. Vermersch, P.-O. Guimond, H. Pichler, and P. Zoller, Quantum State Transfer via Noisy Photonic and Phononic Waveguides, Phys. Rev. Lett. \textbf{118}, 133601 (2017).
\bibitem{WK2020}W.-K. Mok, D. Aghamalyan, J.-B. You, T. Haug, W. Zhang, C. E. Png, and L.-C. Kwek, Long-distance dissipation-assisted transport of entangled states via a chiral waveguide, Phys. Rev. Research \textbf{2}, 013369 (2020).

\bibitem{YX2025} Y. He, and Y.-X. Zhang, Quantum State Transfer via a Multimode Resonator, Phys. Rev. Lett. \textbf{134}, 023602 (2025).

\bibitem{ZL2023}Z.-L. Xiang, D. G. Olivares, J. J. Garc\'{\i}a-Ripoll, and P. Rabl, Universal Time-Dependent Control Scheme for Realizing Arbitrary Linear Bosonic Transformations, Phys. Rev. Lett. \textbf{130}, 050801 (2023).

\bibitem{MJ2006}M. J. Hartmann, F. G. S. L. Brandao, and M. B. Plenio, Strongly interacting polaritons in coupled arrays of cavities, Nat. Phys. \textbf{2}, 849 (2006).
\bibitem{MN2008}M. Notomi, E. Kuramochi, and T. Tanabe, Large-scale arrays of ultrahigh-Q coupled nanocavities, Nat. Photon. \textbf{2}, 741 (2008).
\bibitem{XZ2023}X. Zhang, E. Kim, D. K. Mark, S. Choi, and O. Painter, A superconducting quantum simulator based on a photonicbandgap metamaterial, Science \textbf{379}, 278 (2023).
\bibitem{CH1993}C. H. Bennett, G. Brassard, C. Crepeau, R. Jozsa, A. Peres, and W. K. Wootters, Teleporting an Unknown Quantum State via Dual Classical and Einstein-Podolsky-Rosen Channels, Phys. Rev. Lett. \textbf{70}, 13 (1993).
\bibitem{DC2017}D. Cavalcanti1, P. Skrzypczyk, and I. \v{S}upi\'{c}, All Entangled States can Demonstrate Nonclassical Teleportation, Phys. Rev. Lett. \textbf{119}, 110501 (2017).
\bibitem{YH2019}Y.-H. Luo, H.-S. Zhong, M. Erhard, X.-L. Wang, L.-C. Peng, M. Krenn, X. Jiang, L. Li, N.-L. Liu, C.-Y. Lu, A. Zeilinger, and J.-W. Pan, Quantum Teleportation in High Dimensions, Phys. Rev. Lett. \textbf{123}, 070505 (2019).
\bibitem{BL2022}B. Li , Y. Cao, Y.-H. Li, W.-Q. Cai, W.-Y. Liu, J.-G. Ren, S.-K. Liao, H.-N. Wu, S.-L. Li, L. Li, N.-L. Liu, C.-Y. Lu, J. Yin, Y.-A. Chen, C.-Z. Peng, and J.-W. Pan, Quantum State Transfer over 1200 km Assisted by Prior Distributed Entanglement, Phys. Rev. Lett. \textbf{128}, 170501 (2022).
\bibitem{JP2014}J. Petersen, J. Volz, and A. Rauschenbeutel, Chiral nanophotonic waveguide interface based on spin-orbit interaction of light, Science \textbf{346}, 67 (2014).
\bibitem{RM2014}R. Mitsch, C. Sayrin, B. Albrecht, P. Schneeweiss, and A. Rauschenbeutel, Quantum state-controlled directional spontaneous emission of photons into a nanophotonic waveguide, Nat. Commun. \textbf{5}, 5713 (2014).
\bibitem{PL2017}P. Lodahl, S.Mahmoodian, S. Stobbe, A. Rauschenbeutel, P. Schneeweiss, J. Volz, H. Pichler, and P. Zoller, Chiral quantum optics, Nature \textbf{541}, 473 (2017).
\bibitem{DG2025}D. G. Su\'{a}rez-Forero, M. J. Mehrabad, C. Vega, A. Gonz\'{a}lez-Tudela, and M. Hafezi, Chiral Quantum Optics: Recent Developments and Future Directions, PRX Quantum \textbf{6}, 020101 (2025).
\bibitem{NL2019}N. Leung, Y. Lu, S. Chakram, R. K. Naik, N. Earnest, R. Ma, K. Jacobs, A. N. Cleland, and D. I. Schuster, Deterministic bidirectional communication and remote entanglement generation between superconducting qubits, npj Quantum Inf. \textbf{5}, 18 (2019).
\bibitem{IS2015}I. S\"{o}llner, S. Mahmoodian, S. L. Hansen, L. Midolo, A. Javadi, G. Kir\v{s}ansk\.{e}, T. Pregnolato, H. El-Ella, E. H. Lee, J. D. Song, S. Stobbe, and P. Lodahl, Deterministic photon-emitter coupling in chiral photonic circuits, Nat. Nanotechnol. \textbf{10}, 775 (2015).
\bibitem{SX2021}S. Xiao, S. Wu, X. Xie, J. Yang, W. Wei, S. Shi, F. Song, S. Sun, J. Dang, L. Yang, Y. Wang, Z. Zuo, T. Wang, J. Zhang, and X. Xu, Position-dependent chiral coupling between single quantum dots and cross waveguides, Appl. Phys. Lett. \textbf{118}, 091106 (2021).
\bibitem{DH2022}D. Hallett, A. P. Foster, D. Whittaker, M. S. Skolnick, and L. R.Wilson, Engineering Chiral Light-Matter Interactions in a Waveguide-Coupled Nanocavity, ACS Photonics \textbf{9}, 706 (2022).
\bibitem{MW2025}M. Weng, H. Yu, and Z. Wang, High-fidelity generation of Bell and $W$ states in a giant-atom system via bound states in the continuum, Phys. Rev. A \textbf{111}, 053711 (2025).
\bibitem{AP2021}A. P. M. Place, L. V. H. Rodgers, P. Mundada, B. M. Smitham, M. Fitzpatrick, Z. Leng, A. Premkumar, J. Bryon, A. Vrajitoarea, S. Sussman, G. Cheng, T. Madhavan, H. K. Babla, X. H. Le, Y. Gang, B. J\"{a}ck, A. Gyenis, N. Yao, R. J. Cava, N. P. de Leon, and A. A. Houck, New material platform for superconducting transmon qubits with coherence times exceeding 0.3 milliseconds, Nat. Commun. \textbf{12}, 1779 (2021).
\bibitem{SG2024}S. Ganjam, Y. Wang, Y. Lu, A. Banerjee, C. U Lei, L. Krayzman, K. Kisslinger, C. Zhou, R. Li, Y. Jia, M. Liu, L. Frunzio, and R. J. Schoelkopf, Surpassing millisecond coherence in on chip
superconducting quantum memories by optimizing materials and circuit design, Nat. Commun. \textbf{15}, 3687 (2024).
\bibitem{JB2024}J. Bizn\'{a}rov\'{a}, A. Osman, E. Rehnman, L. Chayanun, C. Kri\v{z}an, P. Malmberg, M. Rommel, C. Warren, P. Delsing, A. Yurgens, J. Bylander, and A. F. Roudsari, Mitigation of interfacial dielectric loss in aluminum-on-silicon superconducting qubits, npj Quantum Inf. \textbf{10}, 78 (2024).
\bibitem{EV2023}E. V. Zikiy, A. I. Ivanov, N. S. Smirnov, D. O. Moskalev, V. I. Polozov, A. R. Matanin, E. I. Malevannaya, V. V. Echeistov, T. G. Konstantinova, and I. A. Rodionov, High-Q trenched aluminum coplanar resonators with an ultrasonic edge microcutting for superconducting quantum devices, Sci. Rep. \textbf{13}, 15536 (2023).
\bibitem{AE2025}A. E. Oriani, F. Zhao, T. Roy, A. Anferov, K. He, A. Agrawal, R. Banerjee, S. Chakram, and D. I. Schuster, Niobium coaxial cavities with internal quality factors exceeding 1.5 billion for circuit quantum electrodynamics, arXiv: 2403. 00286v2 (2025).
\bibitem{YT2025}Y. Tominaga, S. Shirai, Y. Hishida, H. Terai, and A. Noguchi, Intrinsic quality factors approaching 10 million in superconducting planar resonators enabled by spiral geometry, EPJ Quantum Technol. \textbf{12}, 60 (2025).
\bibitem{SM} See the Supplemental Material for details of the analytical and numerical demonstrations of the BICs, the tunneling dynamics in the nonadiabatic state-transfer process, the fidelity for $N_a=4$, and the failure of state transfer when fewer than $N_a$ BICs are present or when BICs are absent.

\bibitem{SL2021}S. Longhi, Rabi oscillations of bound states in the continuum, Opt. Lett. \textbf{46}, 2091 (2021).
\bibitem{KH2023}K. H. Lim, W.-K. Mok, and L.-C. Kwek, Oscillating bound states in non-Markovian photonic lattices, Phys. Rev. A \textbf{107}, 023716 (2023).
\bibitem{AS2023}A. Soro, C. S. Mu\~{n}oz, and A. F. Kockum, Interaction between giant atoms in a one-dimensional structured environment, Phys. Rev. A \textbf{107}, 013710 (2023).
\bibitem{ER2024}E. R. Ingelsten, A. F. Kockum, and A. Soro, Avoiding decoherence with giant atoms in a two-dimensional structured environment, Phys. Rev. Research \textbf{6}, 043222 (2024).
\bibitem{HY2025}H. Yu, X. Zhang, Z. Wang, and J. Wang, Rabi oscillation and fractional population via the bound states in the continuum in a giant-atom waveguide QED setup, Phys. Rev. A \textbf{111}, 053710 (2025).


\end{thebibliography}

\begin{thebibliography}{99}

\bibitem{schur}R. A. Horn and C. R. Johnson, \textit{Matrix Analysis}, 2nd ed. (Cambridge University Press, Cambridge, 2013).

\bibitem{transmon}M. Kjaergaard, M. E. Schwartz, J. Braum\"{u}ller, P.
Krantz, J. I.-J. Wang, S. Gustavsson, and W. D. Oliver,
Superconducting Qubits: Current State of Play, Annu.
Rev. Condens. Matter Phys. {\bf 11}, 369 (2020).
\end{thebibliography}
\end{document}